\documentclass[10pt,a4paper]{article}
\usepackage{dcolumn}
\usepackage{bm}
\usepackage{amsmath,amssymb}
\usepackage{mathtools}
\usepackage{subcaption}
\usepackage{booktabs}
\usepackage{graphicx,psfrag}
\usepackage{epsfig}
\usepackage{color} 
\usepackage{rotating}
\usepackage{slashed}
\usepackage{geometry}
\usepackage{rotating}
\usepackage{breqn}
\usepackage{caption}
\usepackage{pstricks}
\usepackage{color}
\usepackage{float}
\usepackage{subcaption}
\usepackage{epsfig}
\usepackage{cite}
\usepackage[T1]{fontenc}
\usepackage[utf8]{inputenc}
\allowdisplaybreaks[3]
\numberwithin{equation}{section} 
\def\gsim{\lower0.5ex\hbox{$\:\buildrel >\over\sim\:$}}
\def\lsim{\lower0.5ex\hbox{$\:\buildrel <\over\sim\:$}}
\newcommand{\comment}[1]{}
\newcommand{\beq}{\begin{equation}}
\newcommand{\eeq}{\end{equation}}
\newcommand{\bea}{\begin{eqnarray}}
\newcommand{\eea}{\end{eqnarray}}
\newcommand{\barr}{\begin{array}}
\newcommand{\earr}{\end{array}}
\begin{document}
%\noindent  

\begin{center}

%\title
{\sc \Large \bf A comparative study of the $S_1$ and $U_1$ leptoquark effects in the light quark regime}

\vspace{5mm}
% authors
{\sc  Ilja Dor\v{s}ner 
$^{a}$ 
, Svjetlana Fajfer
$^{b,c}$
and Monalisa Patra
$^{b}$
}
\vspace{5mm}

{\sl \small
$^{a}$  
University of Split, Faculty of Electrical Engineering,
Mechanical Engineering and Naval Architecture in Split (FESB), Ru\dj era Bo\v{s}kovi\'ca 32, 21000 Split, Croatia
}\\
{\sl \small
$^{b}$  
Jo\v{z}ef Stefan Institute, Jamova 39, P.O.\ Box 3000, 1001 Ljubljana, Slovenia 
}\\
{\sl \small
$^{c}$ 
Department of Physics, University of Ljubljana, Jadranska 19, 1000 Ljubljana, Slovenia}
\end{center}

 % Abstract
\vspace*{0.5cm}
\noindent

We study the phenomenology of two leptoquarks, the Standard Model $SU(2)$ singlets $S_1$ and $U_1$, with regard to the latest experimental data from the low-energy flavor physics measurements, LHC, and the IceCube neutrino experiment. We consider a scenario when scalar (vector) leptoquark $S_1$ ($U_1$) couples exclusively to the down quark and the neutrinos (charged leptons) of all flavors, where the leptoquark in question couples to the SM lepton doublets. The couplings of $S_1$ ($U_1$) to the up-type quarks and the charged leptons (neutrinos) are in turn uniquely determined via $SU(2)$ symmetry. We find that the most important constraints on the leptoquark parameter space originate from flavor physics measurements, followed by the LHC search limits that take over the flavor physics ones in the large LQ mass regime. We furthermore show that $S_1$ ($U_1$) marginally improves (spoils) the fit of the current IceCube data with respect to the SM case within the region of parameter space that is otherwise consistent with various low-energy flavor physics measurements and the latest LHC input. Our study offers an up-to-date analysis for these two leptoquarks in view of the latest experimental data. 
% %%%%%%%%%%%%%%%%%%%%%%%%%%%%%%%%%%%%%%%%%%%%%%%%%%
 \setcounter{footnote}{0}
 %\tableofcontents
 %\newpage

\section{Introduction}

The leptoquarks (LQs) are hypothetical particles that directly couple a Standard Model (SM) quark to a lepton. There are 12 (10) types of multiplets~\cite{Dorsner:2016wpm} under the SM gauge group $SU(3)\times SU(2)\times U(1)$ with this ability if one assumes presence (absence) of the right-handed neutrinos. They can either be of scalar or vector nature but are, in all instances, triplets under the SM $SU(3)$ group. The LQs emerge in a natural way in many New Physics (NP) proposals such as the Grand Unified Theories based on Pati-Salam model~\cite{Pati:1973uk,Pati:1974yy}, $SU(5)$~\cite{Dorsner:2012uz}, $SO(10)$~\cite{Georgi:1974sy,Gershtein:1999gp}, supersymmetry with R-parity violation~\cite{Barbier:2004ez}, and composite models~\cite{Schrempp:1984nj,Wudka:1985ef}. Most recently LQs have been singled out as the most promising candidates for the explanations of anomalies in the low-energy flavor physics experiments~\cite{Aaij:2015yra,Aaij:2017vbb,Aaij:2014ora,Aaij:2019wad} concerning the $B$ meson semileptonic decays that hint at the lepton flavor universality violation. The anomalies in question, i.e., $R_{D^{(\ast)}}$ and $R_{K^{(\ast)}}$, usually require that the LQs couple strongly to the heavy quarks and leptons. This particular regime has thus been studied extensively in the context of both scalar~\cite{Bauer:2015knc,Becirevic:2016yqi,Crivellin:2017zlb,Buttazzo:2017ixm,Azatov:2018kzb,Crivellin:2019qnh} and vector~\cite{Buttazzo:2017ixm,Azatov:2018kzb,Assad:2017iib,DiLuzio:2017vat,Calibbi:2017qbu,Bordone:2017bld,Fornal:2018dqn,Cornella:2019hct} LQs with few notable exceptions~\cite{Bansal:2018eha}. 

We are interested, in this manuscript, in the scenarios when LQs primarily couple to the quarks of the first generation and neutrinos of all flavors and investigate the viability of the associated parameter space spanned by the LQ masses and coupling strengths in view of the latest experimental data from flavor physics, LHC, and the South Pole situated IceCube detector. We accordingly study the implications of the most recent and the most relevant experimental results on the parameter space for two representative LQ scenarios.  One scenario features scalar LQ $S_1$ and the other uses vector LQ $U_1$, where both fields are singlets under the SM $SU(2)$ group, allowing them to couple to both the left- and the right-handed quarks and leptons. We perform, in particular, a thorough analysis of the viability of the $S_1$ ($U_1$) scenario assuming non-zero couplings between $S_1$ ($U_1$), down quark, and neutrinos (charged leptons) of all three generations, where the LQ in question couples to the SM lepton doublets. Consequentially, $S_1$ ($U_1$) couples up-type quarks to charged leptons (neutrinos). We also entertain the possibility that $S_1$ ($U_1$) couples down quark (up quark) to the right-handed neutrinos to investigate the sensitivity of the latest IceCube data to constrain the associated parameter space.

The outline of the paper is as follows. We describe the two LQ scenarios and the flavor ansatz considered in our work in Sec.~\ref{sec:model}. The constraints from the low-energy flavor physics experiments for these two LQ scenarios are presented in Sec.~\ref{sec:low_energy}. The LHC constraints from the single LQ and the LQ pair productions are discussed in Sec.~\ref{sec:LHC_res}. We then perform the data analysis of the IceCube PeV events within these two frameworks in Sec.~\ref{sec:icecube}. The combined analysis using the low-energy flavor observables, along with the LHC results and the latest IceCube data, for both $S_1$ and $U_1$, is presented in Sec.~\ref{sec:combined_analysis}. Finally we conclude in Sec.~\ref{sec:conclusion}. 

\section{Leptoquark scenarios}
\label{sec:model}

We briefly review in this section the LQ scenarios we consider in our work. The two representative scenarios that are addressed in our analyses are the scalar LQ $S_1$ and the vector LQ $U_1$.  

\subsection{Scalar leptoquark $S_1 = ({\bf \bar{3}},{\bf 1},1/3)$}

We study the signatures of $S_1$, whose $SU(3)\times SU(2) \times U(1)$ quantum numbers are $({\bf \bar{3}},{\bf 1},1/3)$, on the flavor, LHC, and IceCube observables. In our normalisation the electric charge of $S_1$ is 1/3 in the absolute units of the electron charge. The relevant Lagrangian terms, in the mass eigenstate basis, are of the form

\begin{eqnarray}\label{lag:S1}
\mathcal{L} &\supset& -(y^{L}U)_{1j} \bar{d}^{C\,1}_L{S}_1\nu_L^j+(V^\ast y^{L})_{ij}\bar{u}^{C\,i}_L{S}_1e^j_L+y^{R}_{1j}\bar{d}^{C\,1}_R{S}_1\nu_R^j
+{\mathrm {h.c.}},
\end{eqnarray}
where the subscripts $i,j(=1,2,3)$ denote the flavor of the quarks and leptons, $V$ is the Cabibbo-Kobayashi-Maskawa (CKM) mixing matrix and $U$ is the Pontecorvo-Maki-Nakagawa-Sakata (PMNS) mixing matrix. We work under the assumption that the only non-zero $S_1$ couplings are $y^L_{11} \equiv y^L_{d\nu_e}$, $y^L_{12} \equiv y^L_{d\nu_\mu}$, and $y^L_{13} \equiv y^L_{d\nu_\tau}$. We also entertain the possibility that $S_1$ couples to the right-handed neutrinos and we set the couplings $y^R_{11} \equiv y^R_{d\nu_e}$, $y^R_{12} \equiv y^R_{d\nu_\mu}$, and $y^R_{13} \equiv y^R_{d\nu_\tau}$ to be equal to each other, if and when switched on. All the other LQ Yukawa couplings are set to zero. Note that the $S_1$ couplings with the up-type quarks and charged leptons are fixed by the CKM mixing matrix.

\subsection{Vector leptoquark $U_1 = ({\bf{3}},{\bf{1}},2/3)$} 

The relevant Lagrangian terms for the $U_1$ LQ, in the mass eigenstate basis, are
\begin{eqnarray}\label{eq:1}
\mathcal{L}\supset(V^\dagger \chi^{L} U)_{ij}\bar{u}^i_L \gamma^\mu U_{1,\mu}\nu_L^j+ \chi_{1j}^{L}\bar{d}^1_L \gamma^\mu U_{1,\mu}e_L^j+ \chi_{1j}^{R}\bar{u}^1_R \gamma^\mu U_{1,\mu}\nu_R^j+{\mathrm {h.c.}}.
\end{eqnarray}
We consider the scenario where $U_1$ only couples to the down quark and charged leptons of all three generations with
\begin{equation}
\chi_{11}^{L} \equiv \chi^L_{de}, \quad \chi_{12}^{L} \equiv \chi^L_{d\mu}, \quad \chi_{13}^{L} \equiv \chi^L_{d\tau}.
\end{equation}
We also analyse the possibility when the couplings $\chi_{11}^{R} \equiv \chi^R_{u \nu_e}$, $\chi_{12}^{R} \equiv \chi^R_{u \nu_\mu}$, and $\chi_{13}^{R} \equiv \chi^R_{u \nu_\tau}$ of $U_1$ with the up quark and the right-handed neutrinos are switched on and equal to each other. All other $U_1$ couplings are set to zero. 

\section{Low-energy constraints}
\label{sec:low_energy}

The LQ interaction ansatz defined in the previous section can lead to leptonic decays of pseudoscalar mesons or flavor changing processes at both the tree and the one-loop levels. The LQ couplings to the first generation quarks and electron are strongly constrained by the atomic parity violation (APV) experimental results. The experimental upper bounds on the $\ell \rightarrow \ell' \gamma$ decay branching ratios, with the LQ contribution coming in the loop, will also constrain the couplings of $S_1$ with the quarks and leptons. These branching ratios receive contribution from both the left-handed as well as the right-handed couplings of the quarks to the leptons. The upper limits on the lepton flavor violating decays of $\mu$ and $\tau$ leptons are obtained from various experiments with BR$(\mu \rightarrow e \gamma) < 4.2 \times 10^{-13}$~\cite{TheMEG:2016wtm}, BR$(\tau\rightarrow e \gamma) < 3.3 \times 10^{-8}$, and BR$(\tau\rightarrow \mu \gamma) < 4.4 \times 10^{-8}$~\cite{Aubert:2009ag} @ 
90\%\,C.L.. $S_1$ also contributes at the tree level to the rare flavor process $D^0\rightarrow \mu^+\mu^-$. The most recent measurement of this branching ratio comes from LHCb~\cite{Aaij:2013cza} and reads 
${\mathrm{BR}}(D^0\rightarrow \mu^+ \mu^-) < 7.6 \times 10^{-9}$. The same couplings contribute to the $D^0 - \bar D^0$ oscillations. Following the study of the LQ effects in the $D^0 - \bar D^0$ oscillations explained in detail in Refs.~\cite{Dorsner:2016wpm,Fajfer:2015mia}, we require that the LQ contributions are smaller than the current bounds on the $D^0 - \bar D^0$ mixing amplitude.  

Since the LQs, in our case, yield new contributions to $\ell \rightarrow \ell' \gamma$, the APV measurements, the rare meson decays, and the ratio of the leptonic decays of the pseudoscalar meson, we take into account all these constraints. 

\vspace{0.2cm}
 \underline{Lepton flavor violation in the pion sector}
\vspace{0.2cm}

The contribution of weak singlets $S_1$ and $U_1$ to the pion muonic decays is different from the pion electron decays due to the different values of $e$ and $\mu$ couplings with the first generation quarks as well as the dependence on $m_e$ and $m_\mu$. The effects of this type can be exposed by the lepton flavor universality ratios $R^\pi_{e/\mu}$ and $R^\pi_{\tau/\mu}$, where
\begin{eqnarray}\label{eq:flav}
R^\pi_{e/\mu}&=&\frac{{\mathrm {BR}}(\pi^- \rightarrow e^-\bar{\nu})}{{\mathrm {BR}}(\pi^- \rightarrow \mu^-\bar{\nu})}, \quad R^\pi_{\tau/\mu}=\frac{{\mathrm {BR}}(\tau^- \rightarrow \pi^-\bar{\nu})}{{\mathrm {BR}}(\pi^- \rightarrow \mu^-\bar{\nu})},
\end{eqnarray}
with the experimental result $R^\pi_{e/\mu}|_{\mathrm{exp}} =  (1.2327 \pm 0.0023)\times 10^{-4}$ and the SM value $R^\pi_{e/\mu} |_{\mathrm{SM}} = (1.2352 \pm 0.0001)\times 10^{-4}$  \cite{Cirigliano:2007xi} (see Eq.~(\ref{eq:appSM}). The measured ratio is 
$R^\pi_{\tau/\mu}|_{\mathrm{exp}}= 0.1082\pm 0.0005$ \cite{Tanabashi:2018oca}, while the SM value is found to be  $R^\pi_{\tau/\mu}|_{\mathrm{SM}} = 0.1088 \pm 0.0002$, using Eq.~\eqref{eq:appSM}.

\begin{table}[htb!]
\begin{center}
\begin{tabular}{| c | c ||c|c || c|c|} \hline 
 $\tau_D$ (s) & 4.1 $\times 10^{-13}$ & $\tau_\pi$ (s) & 2.603 $\times 10^{-8}$ & $\tau_\tau$ (s) & 2.903 $\times 10^{-13}$ \\ 
 $m_D$ (GeV) & 1.86  & $m_\pi$ (GeV) & 0.140  & $m_\tau$ (GeV) & 1.7768 \\  
 $m_c$ (GeV) & 1.28  & $m_e$ (GeV) & 0.51 $\times 10^{-3}$ & $m_\mu$ (GeV) & 0.105\\
 $f_D$ (MeV) & 212   & $f_\pi$ (MeV) & 130.41  & &\\   \hline 
\end{tabular}
\caption{Numerical values of parameters used in our calculation, taken from PDG~\cite{Tanabashi:2018oca}.}
\label{tab:param}
\end{center}
\end{table}

\begin{figure}[htb]
\centering
\includegraphics[width=16.5cm, height=6.5cm,keepaspectratio]{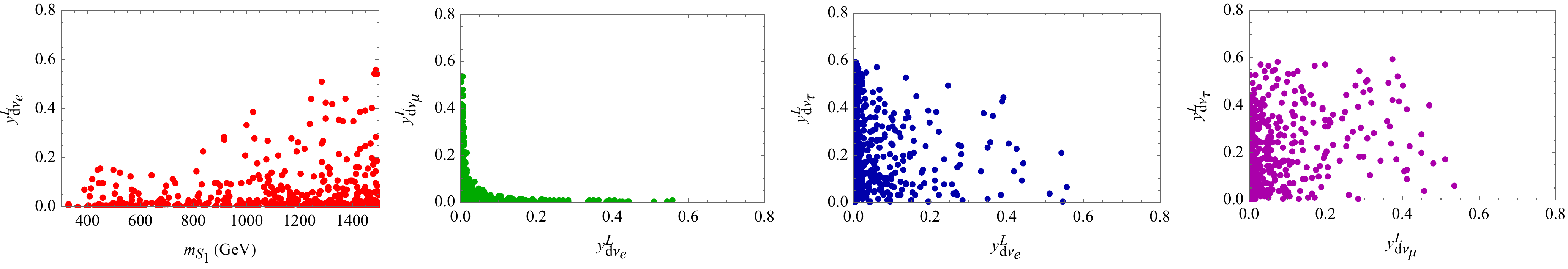}
\includegraphics[width=16.5cm, height=6.5cm,keepaspectratio]{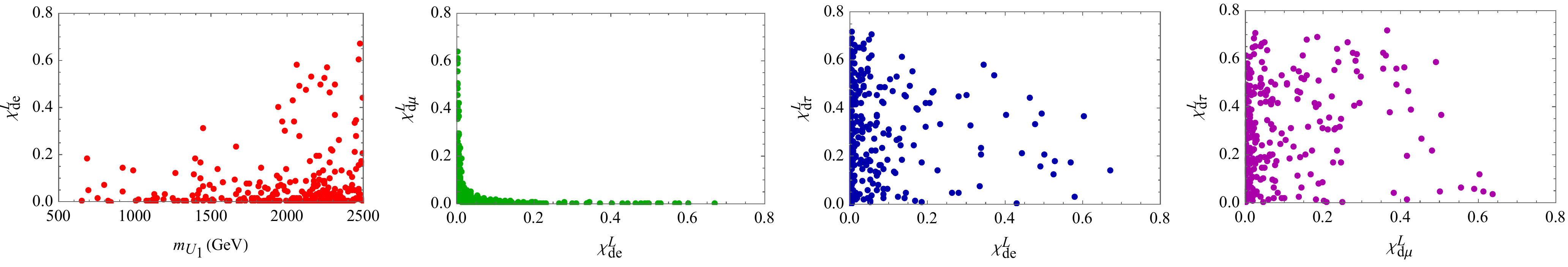}
\caption{The allowed parameter space after taking into account the results from the most relevant low-energy flavor experiments.}
\label{fig:scan_res}
\end{figure}

We list the formulas for the branching ratios of the pion, $D^0$ meson, and the $\tau$ lepton for the $S_1$ and $U_1$ cases in Appendix~\ref{sec:appen} and specify numerical values of input parameters relevant for our analysis in Table~\ref{tab:param}. We furthermore summarize in Fig.~\ref{fig:scan_res} results of a randomized scan within the parameter space ($m_{S_1}\in(300\,{\mathrm {GeV}}, 1.5\,{\mathrm {TeV}})$, $y^L_{d\nu_e},y^L_{d\nu_\mu},y^L_{d\nu_\tau} \in (0.0, 0.8)$) and ($m_{U_1}\in(500\,{\mathrm {GeV}}, 2.5\,{\mathrm {TeV}})$, $\chi^L_{de},\chi^L_{d\mu},\chi^L_{d\tau} \in (0.0, 0.8)$) that takes into account the constraints from the pion sector, APV, the rare radiative decays $\ell \rightarrow \ell' \gamma$, and $D^0\to \mu^+\mu^-$ decays. The plots in Fig.~\ref{fig:scan_res} show currently allowed parameter spaces of the left-handed couplings of the down quark and the first generation leptons as a function of the LQ mass as well as the correlations between different left-handed couplings. We find that $y^L_{d\nu_e}$ and $y^L_{d\nu_\mu}$ ($\chi^L_{de}$ and $\chi^L_{d\mu}$) cannot be simultaneously large due to conflict with the current results from the low-energy sector in the $S_1$ ($U_1$) case. This can be clearly seen in the panels of the second column of Fig.~\ref{fig:scan_res}. We therefore mainly work, in what follows, in the presence of the left-handed coupling of the down quark and the first generation leptons, with the other couplings being set to zero.

The flavor experiments constrain the parameter space of the vector LQ more tightly than that of the scalar one. The $S_1$ LQ also contributes at the loop level to the $Z\rightarrow \ell \bar{\ell}$ decay amplitude, with $S_1$ and the up-type quarks running in the loop. The $Z$ branching ratio to a pair of leptons has been precisely measured at LEP~\cite{Tanabashi:2018oca}, thereby imposing constraints on the $S_1$ parameter space. We have used formula for the one-loop contribution of $S_1$ computed in Ref.~\cite{Arnan:2019olv} and found that the bounds on the LQ couplings from the $Z$ leptonic branching ratio are not up to par with the other experimental constraints considered before. The loop level contribution to the $Z\rightarrow \ell \bar{\ell}$ decay amplitude in case of $U_1$ is also negligible for the parameter space that survives the other low-energy flavor physics experiments. 

We next discuss the LHC limits on the LQ masses and their couplings.

\section{Constraints from the Large Hadron Collider}
 \label{sec:LHC_res}

The LQ couplings to the quark-lepton pairs have been constrained through both direct and indirect searches in a number of collider experiments. Prior to the LHC era, the LQs were searched for at LEP~\cite{Abreu:1993bc}, HERA~\cite{Aid:1995wd,Abramowicz:2012tg} and  Tevatron~\cite{GrossoPilcher:1998qt,Abazov:2008np}. The LQs have been hunted for at the LHC mainly through pair production~\cite{Aaboud:2016qeg,CMS:2016imw,CMS:2016qhm} but there are also several searches/recasts that rely on the single LQ production~\cite{Khachatryan:2015qda} as well as dilepton~\cite{Raj:2016aky} and monolepton~\cite{Bansal:2018eha} Drell-Yan processes to generate constraints. For a summary of currently available bounds on the LQ masses and associated couplings from the LHC searches for various flavor final states see, for example, Refs.~\cite{Schmaltz:2018nls,Bandyopadhyay:2018syt}.

The LQs are pair-produced through either $gg$ or $q\bar{q}$ fusion that is primarily dominated by the QCD interactions. There is also a Yukawa coupling contribution to the LQ pair production, corresponding to a $t$-channel process, with its amplitude being proportional to the product of the two relevant Yukawa couplings. This $t$-channel process is highly suppressed compared to the QCD driven one, unless the Yukawa couplings are rather large. The production cross section at the LHC for a pair of vector LQs, when compared to the scalar ones, additionally depends on the underlying theory for the origin of vector LQs. The relevant trilinear and quartic couplings of vector LQs to a gluon or a pair of gluons is then completely fixed by the extended gauge invariance of the model wherein the vector LQs appear as fundamental objects. We here work in the limit, where the vector LQ is some low energy manifestation of a more fundamental theory at high energy scale. The vector LQ-gluon interaction terms can then be obtained from the most general $SU(3)$ invariant effective Lagrangian given by
\begin{equation}
    \mathcal{L}_V \supset -i g_s \kappa U_{1,\mu}^\dagger G^{\mu\nu}U_{1,\nu},
    \end{equation}
where $G^{\mu\nu}$ is the gluon field strength tensor and $\kappa$ is a dimensionless parameter which we consider to be 1 for our calculations. 
    
Dedicated studies have been performed at the LHC, assuming LQ pair production and a 100\% branching ratio (BR) of LQ decaying into a charged lepton and jet ($jj\ell^+\ell^-$, where $\ell = e,\mu$) or to jet and missing energy ($jj\nu\nu$). Considering the model Lagrangian in Eq.~\eqref{lag:S1} (Eq.~\eqref{eq:1}) in the $S_1$ ($U_1$) case, the final states relevant for our analysis are $jjee$, $jj\mu\mu$, and $jj\nu\nu$, with $j\nu$ having the dominant branching ratio. There being no distinction made in the LHC LQ searches in case of the light quark jets ($u,d,c,s$), we can consider the LHC limit directly. The upper limits on the LQ production cross section times BR$^2$ for these final states are provided by the LHC collaborations~\cite{Sirunyan:2018kzh,CMS:2018bhq,Aaboud:2016qeg}. The LQs coupling to first-generation quarks and electrons or muons are also sought in single production processes, i.e., $pp \rightarrow \ell^+ \ell^- j$~\cite{Khachatryan:2015qda}. This process occurs via $s$- and $t$- channel quark-gluon fusion and is directly proportional to the Yukawa coupling of the $u(d)$ quarks to the leptons. The single production of LQs at 8\,TeV LHC~\cite{Khachatryan:2015qda} is also considered in our analysis.  
\begin{figure}[htb!]
\centering
\begin{subfigure}{.45\textwidth}
\includegraphics[width=0.9\linewidth, height=0.2\textheight, keepaspectratio]{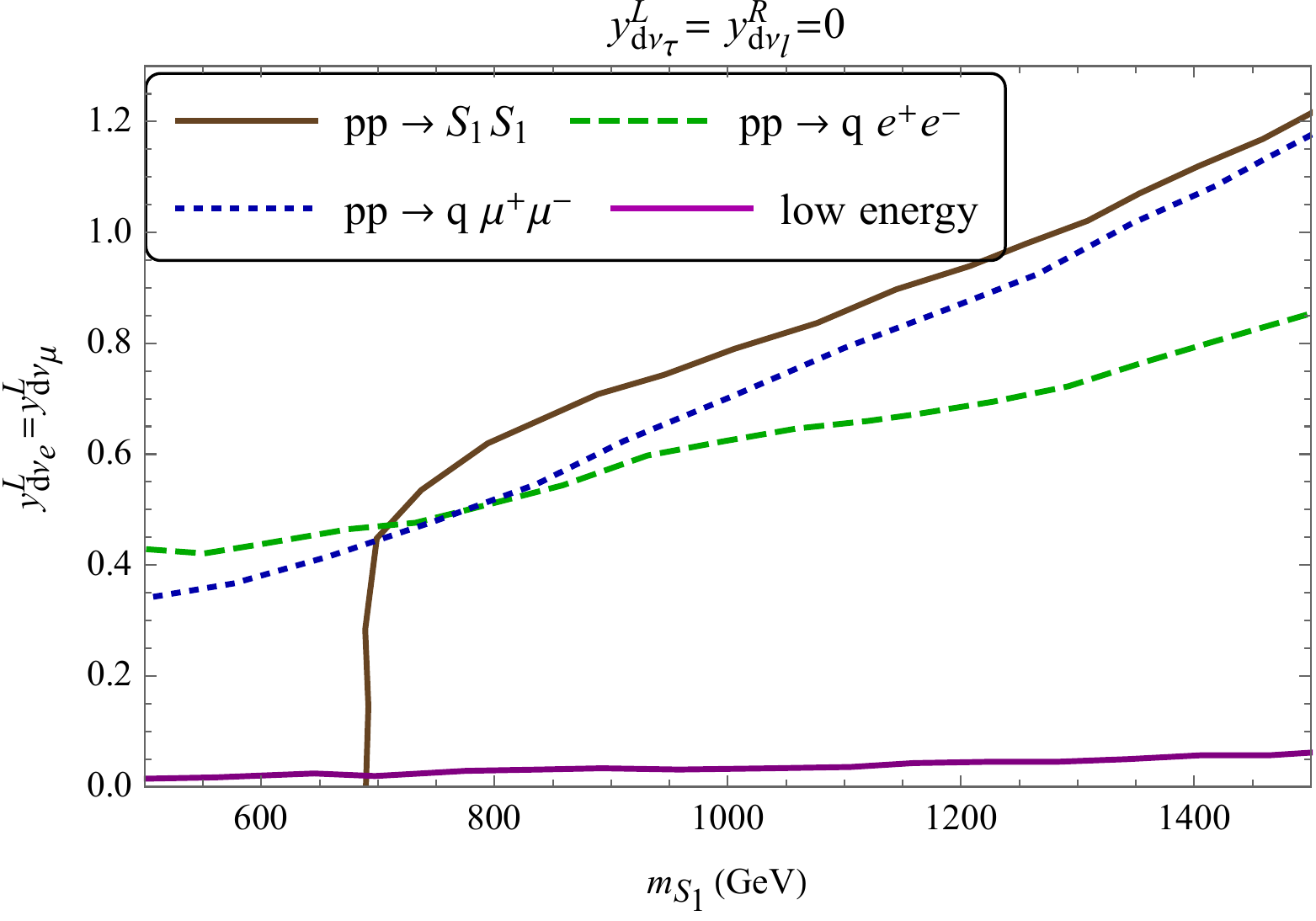}
\caption{}
\label{fig:s1}
\end{subfigure}
\begin{subfigure}{.45\textwidth}
\includegraphics[width=0.9\linewidth, height=0.2\textheight, keepaspectratio]{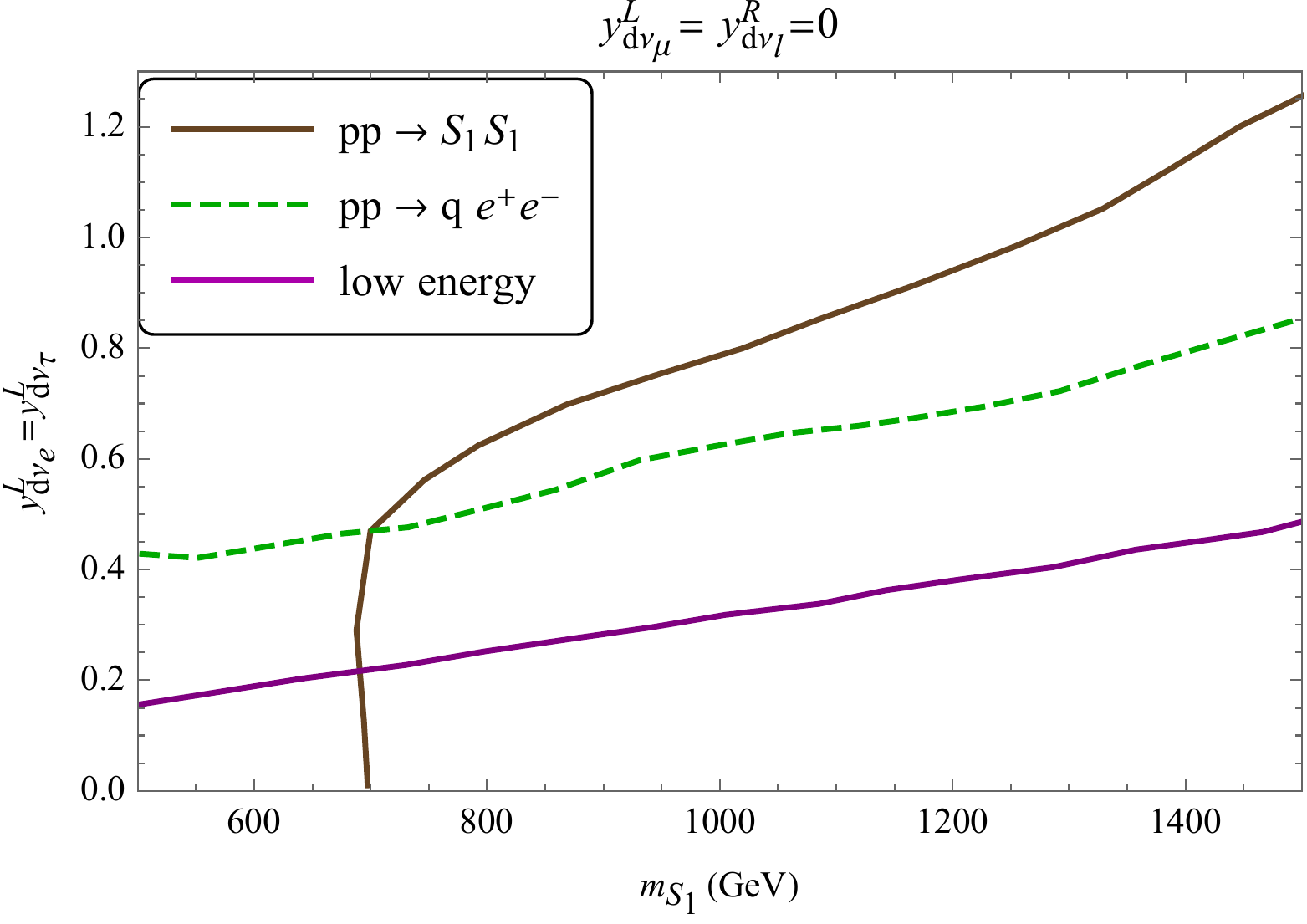}
\caption{}
\label{fig:s2}
\end{subfigure}
\begin{subfigure}{.45\textwidth}
\includegraphics[width=0.9\linewidth, height=0.2\textheight, keepaspectratio]{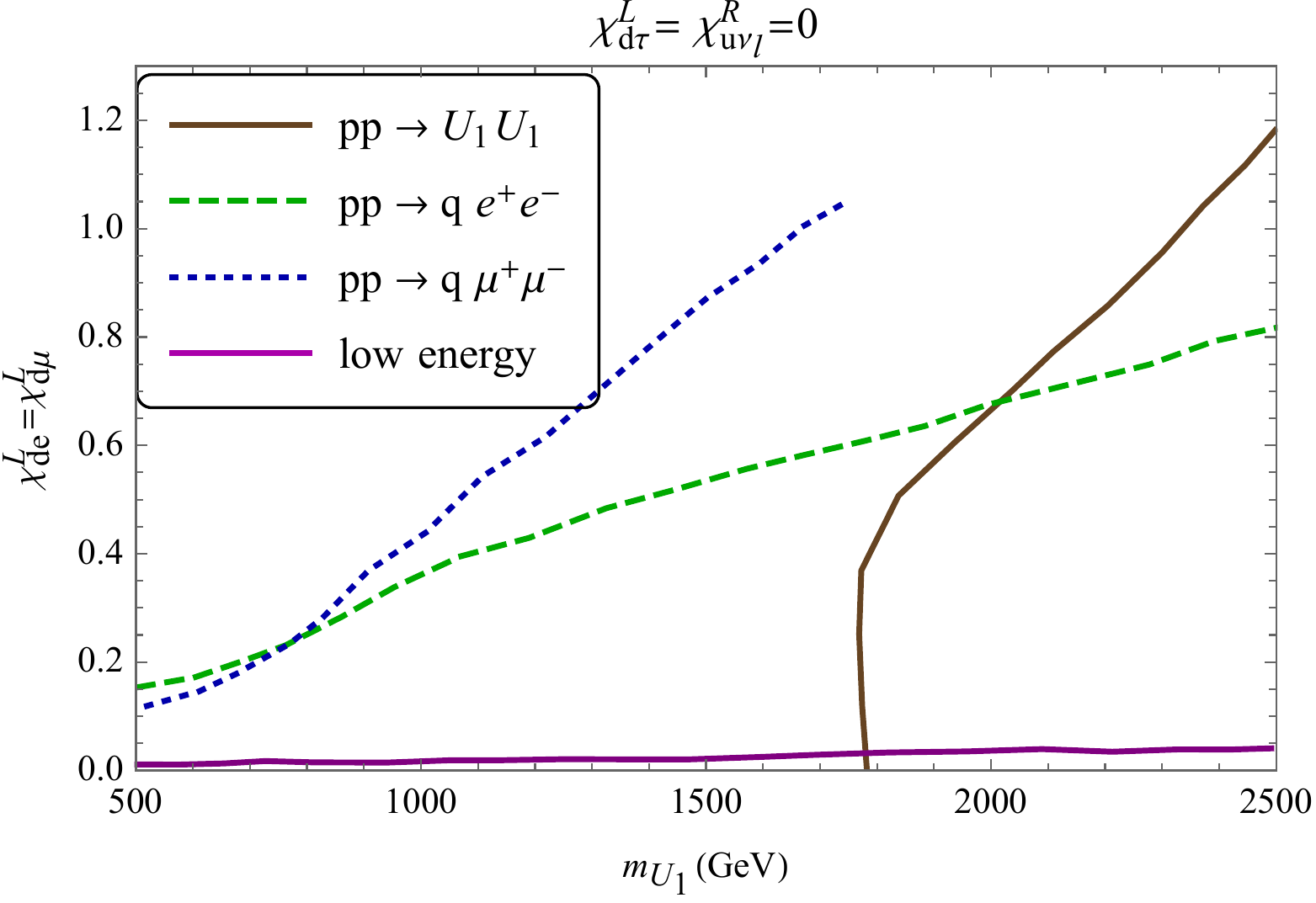}
\caption{}
\label{fig:s3}
\end{subfigure}
\begin{subfigure}{.45\textwidth}
\includegraphics[width=0.9\linewidth, height=0.2\textheight, keepaspectratio]{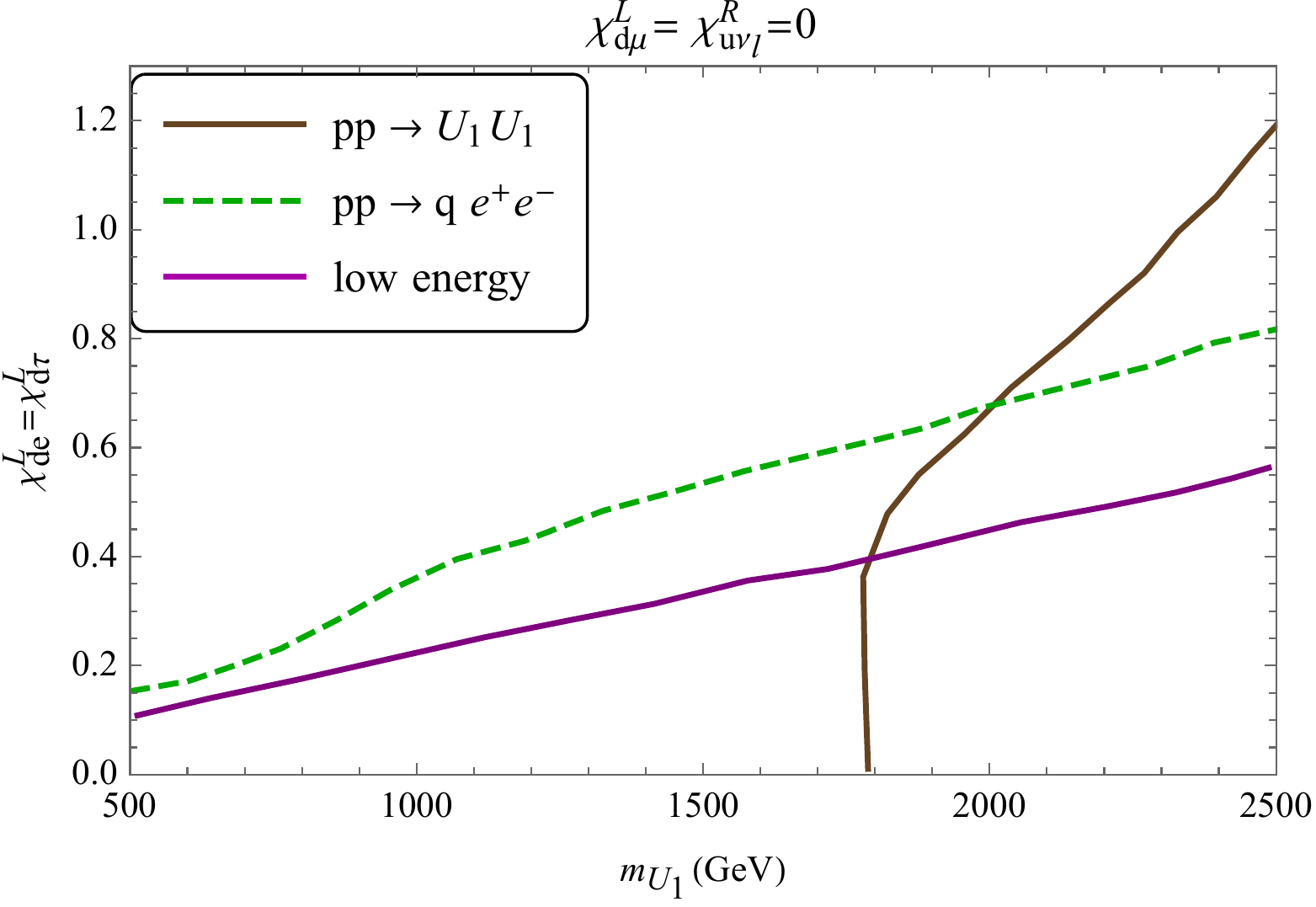}
\caption{}
\label{fig:s4}
\end{subfigure}
\caption{The allowed parameter space in case of $S_1$ and $U_1$ from different experiments discussed in the text. The area below the brown (blue and green) line is compatible with the LQ searches through pair (single LQ) production, at the 95\%\,C.L., at the $13$\,TeV ($8$\,TeV) LHC. The region enclosed by the purple line is allowed by the low-energy flavor experiments, at the 2\,$\sigma$ level.}
\label{fig:lhc_ic_le}
\end{figure}

We show in Fig.~\ref{fig:lhc_ic_le} the allowed parameter space for two different choices of values of relevant couplings of $S_1$ ($U_1$) as a function of $m_{S_1}$ ($m_{U_1}$). In an effort to confront the LHC constraints with the flavor physics measurements, we consider two particular scenarios: ($i$) $y^L_{d\nu_e}(\chi^L_{de})$ = $y^L_{d\nu_\mu}(\chi^L_{d\mu})$ and ($ii$) $y^L_{d\nu_e}(\chi^L_{de})$ = $y^L_{d\nu_\tau}(\chi^L_{d\tau})$. The region below the purple line in Fig.~\ref{fig:lhc_ic_le} is currently allowed by the low-energy experiments at the 2\,$\sigma$ level, as discussed in Sec.~\ref{sec:low_energy}. The parameter space allowed by the LHC, at 95\% C.L., is the region below the brown line in case of LQ pair production and the subsequent decay to light quarks and leptons. The allowed region from the single LQ production is the area below the green (for $qe^+e^-$) and blue (for $q \mu^+\mu^-$) lines.  In the single LQ production, the process $pp\rightarrow q e^+e^-$ is proportional to the Yukawa coupling $y^L_{d\nu_e}$ ($\chi^L_{de}$), whereas $pp\rightarrow q \mu^+\mu^-$ is proportional to $y^L_{d\nu_\mu}$ ($\chi^L_{d\mu}$) for the $S_1$ ($U_1$) case. Therefore no limit is obtained from the process $pp\rightarrow q \mu^+\mu^-$, in case of $y^L_{d\nu_e}(\chi^L_{de})$ = $y^L_{d\nu_\tau}(\chi^L_{d\tau})$.

The low-energy experiments, as discussed in the previous section, do not allow simultaneous presence of large values for the couplings of the first generation quarks to electrons and muons. Therefore the choice $y^L_{d\nu_e}(\chi^L_{de})$ = $y^L_{d\nu_\mu}(\chi^L_{d\mu})$ is strongly constrained by the flavor physics measurements. The single LQ production has more stringent limits than the pair production for $ee$ final state at large LQ masses. This is due to the final state phase space. The blue dotted line in Fig.~\ref{fig:s3} is incomplete as the experimental result from $pp\rightarrow q \mu^+\mu^-$ is provided up to the LQ mass of $1.8$\,TeV. The most stringent constraint on the available parameter space currently comes from the low-energy flavor experiments denoted by the purple line. The LHC direct searches from the LQ pair production currently exclude $m_{S_1} < 700$\,GeV, and $m_{U_1} < 1700$\,GeV, irrespective of the choice of couplings as long as LQ decays promptly.  A study done in Ref.~\cite{Bansal:2018eha} has shown that stringent limits can also be obtained on the strength of the LQ coupling to the first generation quarks and left-handed electrons and muons through the monolepton searches. 

We study next in details the effects of $S_1$ and $U_1$ on the observed IceCube PeV events. 

\section{PeV events in IceCube}
\label{sec:icecube}
The observation of the High Energy Starting Events above 100\,TeV  at the South Pole situated IceCube detector~\cite{Aartsen:2013jdh, Aartsen:2014gkd, Aartsen:2017mau, Schneider:2019ayi}, consistent with a flux of high-energy astrophysical neutrinos from outside the galaxy, has motivated a large number of studies that explore the IceCube potential to test various NP models. Since some of the most studied NP sources are various LQ scenarios our intention is to investigate whether the latest IceCube data~\cite{Schneider:2019ayi} might offer an independent way to constrain the LQ mass $m_{\mathrm{LQ}}$ and the strength of its couplings to the quark-lepton pairs that would be on par with the flavor and collider physics constraints.

The high-energy neutrinos coming from outside the atmosphere are detected in the IceCube detector by observing the Cherenkov light emitted by the secondary charged particles produced in the interaction of the neutrinos with the nucleus present in the ice. The charged current (CC) and the neutral current (NC) interactions have distinctive topologies depending on the flavor of the incoming neutrinos. The shower-like events are induced by CC of $\nu_e$ and $\nu_\tau$ interactions and NC interactions of neutrinos of all flavors. The tracks are produced in the CC interactions of $\nu_\mu$ and $\nu_\tau$ ($\tau$ produced in the final state decays to $\nu_\tau \nu_\mu \mu$, giving a distinctive double cascade signature). The expected total number of events at the IceCube from the NC or CC interactions in the deposited energy interval [$E^i_{\mathrm{dep}},E^f_{\mathrm{dep}}$] can be written as
\begin{equation}\label{eq:noofevents}
\mathcal{N}^{k,ch,f}_{\nu_\ell}=T~N_A \int_{E^i_{\mathrm{dep}}}^{E^f_{\mathrm{dep}}} dE_{\mathrm{dep}}\int_0^{\infty}dE_\nu ~Att^f_{\nu_\ell}(E_\nu)  \frac{d\phi^f_{\nu_\ell}}{dE_\nu} \int_0^1 dy M_{\mathrm{eff}}(E^{k,ch}_{\mathrm{true}}) R(E^{k,ch}_{\mathrm{true}},E_{\mathrm{dep}},\sigma(E^{k,ch}_{\mathrm{true}})) \frac{d\sigma^{ch}_{\nu_\ell}(E_\nu,y)}{dy},
\end{equation}
where $T$ is the exposure time in seconds comprising 2635 days of data taking~\cite{Schneider:2019ayi}, $N_A$ is the Avogadro's number 6.022$\times 10^{23}$, $k$ is showers and tracks for each channel $\nu_\ell = \lbrace e,\mu,\tau \rbrace$ induced by the charged and neutral current interactions ($ch$) for an incoming neutrino flux of type $f$ (astrophysical ($a$), conventional atmospheric ($\nu$) or prompt atmospheric flux ($p$)). The effective mass of the detector, $M_{\mathrm{eff}}(E^{k,ch}_{\mathrm{true}})$, is a function of the true electromagnetic equivalent energy and is defined as the mass of the target material times the efficiency of converting the true deposited energy of the event into an observed signal. The energy resolution function is given by $R(E^{k,ch}_{\mathrm{true}},E_{\mathrm{dep}},\sigma(E^{k,ch}_{\mathrm{true}}))$ and is represented by a Gaussian distribution~\cite{Palomares-Ruiz:2015mka}. The effect of the earth's attenuation, in case of neutrino's energy above a few TeV, where the mean free path inside the earth becomes comparable to the distance travelled by the neutrino, is denoted by $Att_{\nu_\ell}(E_\nu)$. The incoming neutrino flux is given by $d\phi^f_{\nu_\ell}/dE_\nu$, where the incoming astrophysical neutrino flux follows the isotropic single unbroken power-law spectrum. This spectrum is given by~\cite{Schneider:2019ayi}
\begin{equation}
\frac{d\phi^{\mathrm{astro}}_{\nu_\ell}}{dE_\nu}  = 3 \Phi_0 f_\ell\Bigg(\frac{E_\nu}{100\,\mathrm{TeV}}\Bigg)^{-\gamma},
\end{equation}
where $f_\ell$ is the fraction of neutrinos of each flavor $\ell$. The fit is performed assuming a $(1/3:1/3:1/3)_\oplus$ flavor ratio, which yields the best fit value for the spectral index $\gamma  = 2.89^{+0.20}_{-0.19}$, with a normalization $\Phi_0 = 6.45^{+1.46}_{-0.46}\times 10^{-18}$\,GeV cm$^{-2}$s$^{-1}$sr$^{-1}$ at 1$\sigma$ significance. 

The neutrino-nucleon differential cross section for different channels in case of the CC and the NC interactions is given by $d\sigma^{ch}_{\nu_\ell}(E_\nu,y)/dy$. The SM differential cross section is given by Eq.~\eqref{eq:totCS_SM} in Appendix~\ref{sec:2a}. At the IceCube detector the neutrinos interact with the nucleons present in the ice. We assume that the natural ice nucleus can be treated as an isoscalar with 10 protons and 8 neutrons. We calculate the event spectra of showers and tracks for each flavor in case of SM assuming an isotropic power-law spectrum. Since we find that the largest contribution to the event spectra comes from the $\nu_e$ showers, the electron neutrino should be sensitive to the NP effects if one is to have an enhanced effect compared to the SM. We therefore study next the effect of $S_1$ and $U_1$ on the IceCube spectrum when these LQs couple the first generation quark to the electron.

\begin{figure}[htb]
\centering
\includegraphics[width=15.5cm, height=3.5cm]{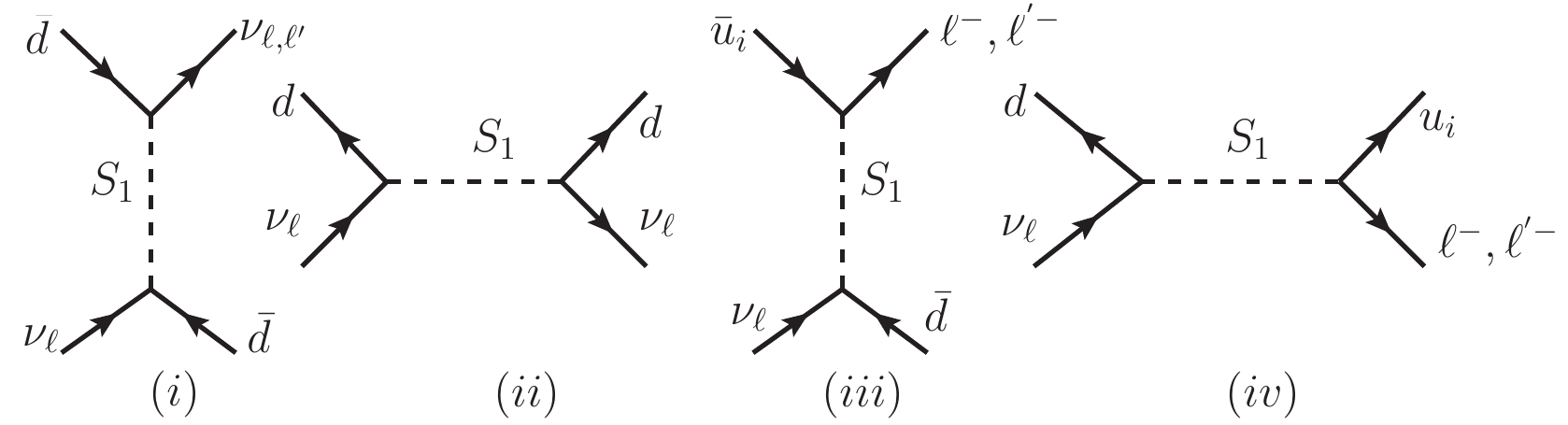}
\caption{The relevant Feynman diagrams for the neutrino-quark interactions mediated by the scalar LQ $S_1$.}
\label{fig:fd_s1}
\end{figure}

The scalar LQ $S_1$ mediates the NC interactions $\nu_\ell d \rightarrow \nu_{\ell} d$ and $\nu_\ell d \rightarrow \nu_{\ell'} d$ and the CC interactions $\nu_\ell \bar{u} \rightarrow \ell\bar{d}$, $\nu_\ell \bar{u} \rightarrow \ell'\bar{d}$, and $d\nu_\ell \rightarrow u\ell$, $d\nu_\ell \rightarrow u\ell'$, where $\ell \neq \ell'$. The Feynman diagrams for the relevant processes are shown in Fig.~\ref{fig:fd_s1}. The charm contribution towards the $t$-channel CC process depicted in Fig.~\ref{fig:fd_s1}($iii$), due to small PDFs, is maximally around $0.001\%$ for the choice of the mass and the couplings considered here and is therefore neglected. The differential $\nu_j N$ cross sections, in the presence of the $S_1$ interactions, are given by Eq.~\eqref{eq:s1_cs} in Appendix~\ref{app_2a}.

\begin{figure}[htb]
\centering
\includegraphics[width=15.5cm, height=3.5cm]{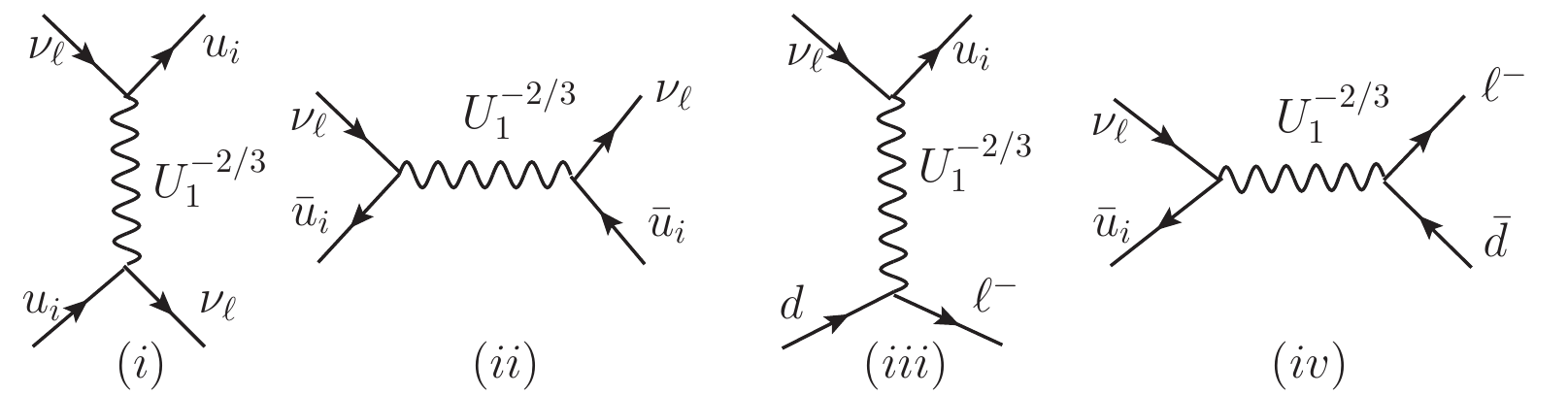}
\caption{The relevant Feynman diagrams for the neutrino-quark interactions mediated by the vector LQ $U_1$.}
\label{fig:fd_u1}
\end{figure}

\begin{figure}[!htb]
\centering
\includegraphics[width=1.0\linewidth, height=0.2\textheight,keepaspectratio]{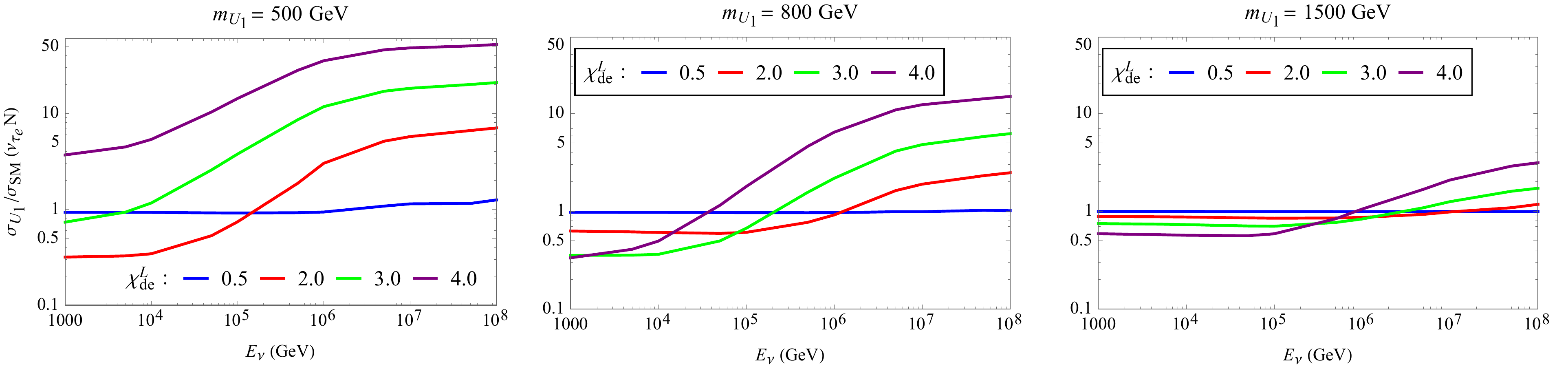}
\caption{The ratio of $\nu_e N$ total cross sections between the  SM + $U_1$ and the SM for different masses and couplings of $U_1$.}
\label{fig:cs_u1}
\end{figure}

The vector LQ $U_1$ contributes to both the NC and CC interactions. The relevant Feynman diagrams, in the presence of $U_1$, are shown in Fig.~\ref{fig:fd_u1}. The $u_i$ in Fig.~\ref{fig:fd_u1} represents the contributions from all three generations of up-type quarks. Note, however, that the charm contributions towards the NC processes, due to small PDFs, are of the order of 0.001\% and can be safely neglected. The differential $\nu_j N$ cross sections in the presence of the $U_1$ interaction are given by Eq.~\eqref{eq:VLQQQ0bar1} in Appendix~\eqref{app_2b}. The $U_1$ LQ compared to the $S_1$ case interferes with the SM leading to interesting features.
 Since the $U_1$ LQ interferes with the SM contribution, we show in Fig.~\ref{fig:cs_u1} the ratio of the $\nu_e N$ total cross sections for the SM + $U_1$ and the SM for different values of masses and couplings. The interference effect is clearly visible for low values of mass and large values of $\chi^L_{de}$. The seven years of IceCube data have fewer events when compared to the SM in the $200$--$300$\,TeV energy range whereas for energies above $1000$\,TeV there are more events when compared to the SM. It can be seen from Fig.~\ref{fig:cs_u1} that there is a crossover in the relevant energy range making it an interesting feature for a more detailed study. We would like to point out that the inclusion of $\chi^L_{d\mu}$ and/or $\chi^L_{d\tau}$ will push the crossover away from the interesting energy range. The three couplings then have to be adjusted so as to get the required effect. 

We study whether the SM + LQ scenarios result in a better or a worse fit of the IceCube data compared to the SM case by calculating parameter $\delta$ that corresponds to the percent change in $\chi^2$~\cite{Chauhan:2017ndd}. We accordingly define
\begin{equation}\label{eq:chi_sq}
\chi^2_{\mathrm{model}} = \sum_{i\geq 100\,\mathrm{TeV}}^{\mathrm  {bins}}\frac{(N_{\mathrm{model}_i}-N_{\mathrm{data}_i})^2}{N_{\mathrm{data}_i}},\quad \quad \delta = 100\% \times \frac{\chi^2_{\mathrm{SM}}-\chi^2_{\mathrm{SM + LQ}}}{\chi^2_{\mathrm{SM}}},
\end{equation}
where the observed number of events $N_{\mathrm{data}_i}$ in each bin $i$ is compared with the LQ scenario prediction and, in our case, $\mathrm{LQ}=S_1,U_1$. We consider the events in the neutrino deposited energy range $[60\,\mathrm{TeV}, 10\,\mathrm{PeV}]$ that is divided in 20 logarithmic energy bins. We initially use only the data for the bins with the non-zero number of events. 

\begin{figure}[!htb]
\begin{subfigure}{.46\textwidth}
\includegraphics[width=7.6cm, height=7.6cm,keepaspectratio]{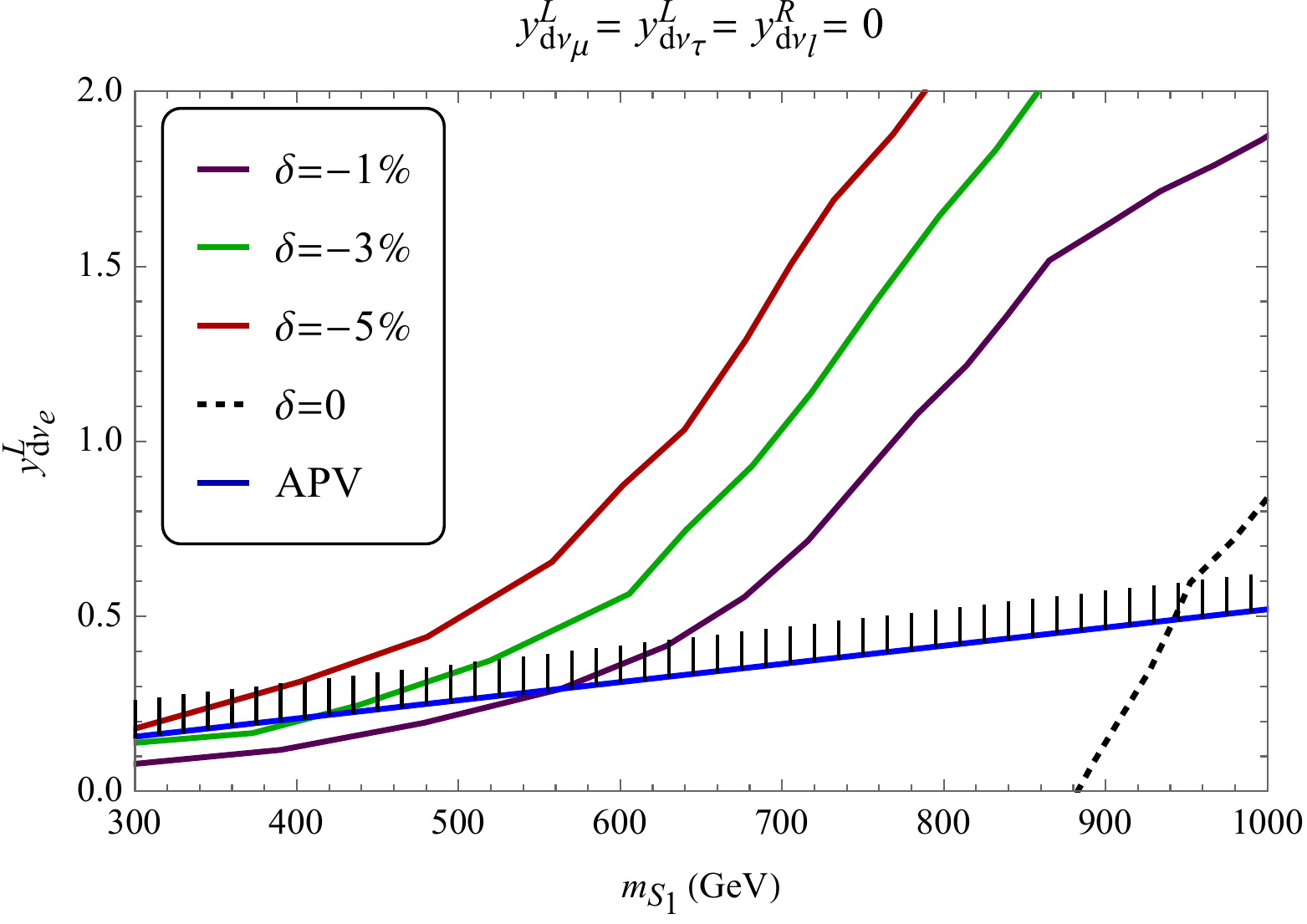}
\caption{}
\label{fig:elec_exclusion}
\end{subfigure}
\begin{subfigure}{.46\textwidth}
 \includegraphics[width=7.6cm, height=7.6cm,keepaspectratio]{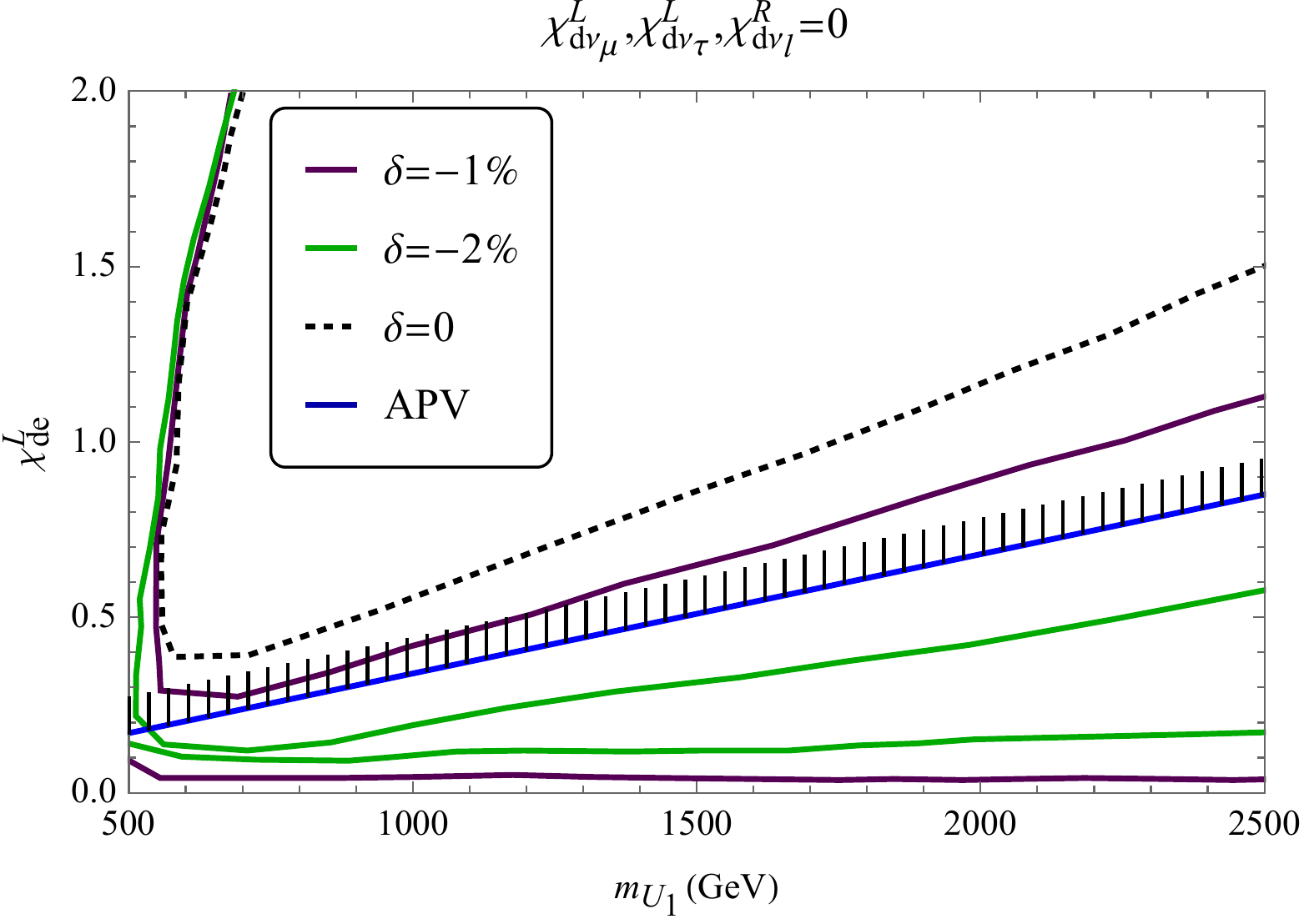}
 \caption{}
  \label{fig:VLQ2}
\end{subfigure}
  \caption{The contour lines of constant $\delta$ in the $m_{S_1(U_1)}$-$y^L_{d\nu_e}(\chi^L_{de})$ plane showing the percent change of $\chi^2$ compared to the SM case. The region to the right of the dotted curve in case of $S_1$ and above the dotted curve in case of $U_1$ results in a fit comparable or better than the SM one. The region above the blue line is  excluded at the 2\,$\sigma$ level by the limits from the APV experiments.}
  \label{fig:elec_exclusion2}
\end{figure}

The SM value with the best fit value of $\gamma$ and $C_0$ from the IceCube data results in a $\chi^2$ value of 0.15. This shows that the current IceCube data is quite compatible with the SM. The NP contribution to the number of events in each bin depends on the values of $y^L_{d\nu_e,d\nu_\mu,d\nu_\tau}$, $y^R_{d\nu_\ell}$, and $m_{S_1}$ ($\chi^L_{d e,d \mu,d \tau}$, $\chi^R_{u \nu_\ell}$, and $m_{U_1}$) in the $S_1$ ($U_1$) case. 
We present in Fig.~\ref{fig:elec_exclusion} contours of constant $\delta$ for the SM + $S_1$ scenario in the $m_{S_1}$-$y^L_{d\nu_e}$ plane. Since the $S_1$ contribution simply adds to the SM one, a small mass and a large value for the LQ-neutrino-quark coupling will lead to an enhanced number of events in each bin. This is beneficial for (detrimental to) the bins where there is an observed excess (lack) of events compared to the SM case. The hatched region above the blue line in Fig.~\ref{fig:elec_exclusion} is currently excluded at the 2\,$\sigma$ level by the APV results. The region to the right of the black dotted curve in Fig.~\ref{fig:elec_exclusion} results in a fit to the IceCube data that is marginally better than the SM one.  

The contours of constant value of $\delta$ in the SM + $U_1$ scenario are shown in Fig.~\ref{fig:VLQ2} for the $m_{U_1}$-$\chi^L_{de}$ parameter space. The region above the black dotted curve results in a fit of the IceCube data that is better than the SM one. It can be seen, through comparison of Figs.~\ref{fig:elec_exclusion} and \ref{fig:VLQ2}, that the $U_1$ scenario has much larger region of parameter space that results in a better fit compared to the SM case than the $S_1$ scenario due to the fact that $U_1$ signatures interfere with the SM. The 2\,$\sigma$ limit on $\chi^L_{de}$ from the APV experiment, i.e., $\chi^L_{de} \leq 0.34 \times m_{U_1}/(1\,\mathrm{TeV}$), as a function of $m_{U_1}$ is shown by the blue line, with the region above the blue line being excluded.

\begin{figure}[!htb]
\centering
\includegraphics[width=0.45\linewidth,height=0.27\textheight,keepaspectratio]{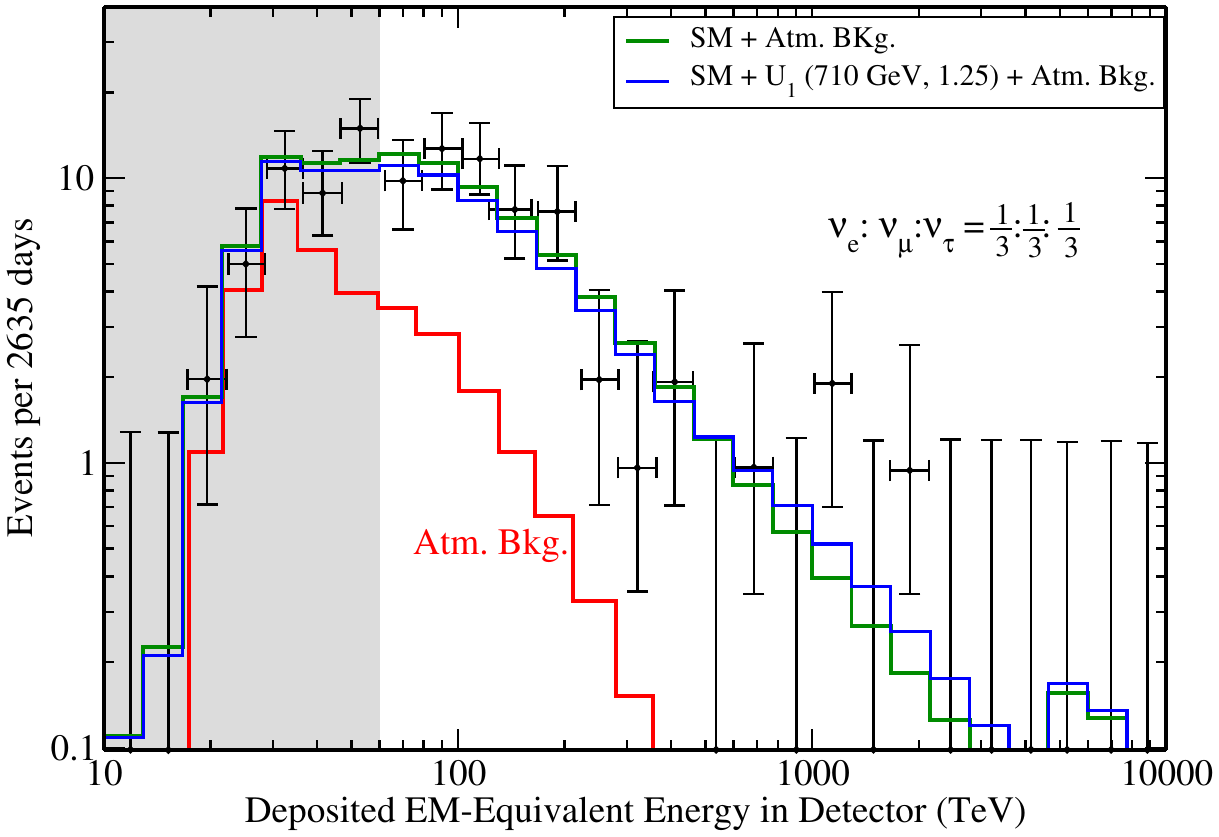}
\includegraphics[width=0.45\linewidth,height=0.27\textheight,keepaspectratio]{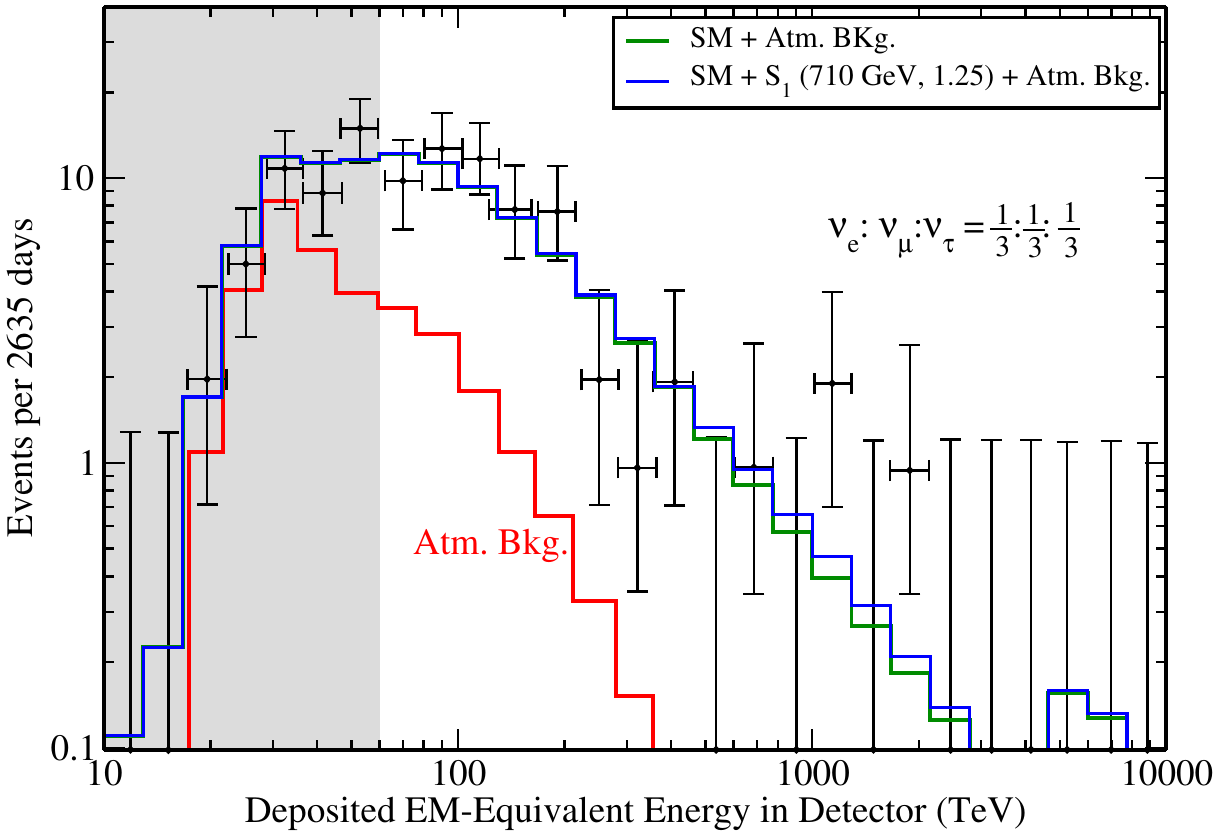}
\caption{The total event rate, with the LQ contribution for $m_{U_1} = 710$\,GeV ($\chi^L_{de} = 1.25$, $\chi^L_{d\mu}, \chi^L_{d \tau} = 0$) (left panel), and $m_{S_1} = 710$\,GeV ($y^L_{d\nu_e} = 1.2$, $y^L_{d\nu_\mu} = y^L_{d\nu_\tau} = 0$) (right panel). The gray shaded region and the bin with zero events are not included in the fit.}
\label{fig:VLQ_S1}
\end{figure}

The couplings of $S_1$ and $U_1$ with the right-handed neutrinos are first fixed to zero, for simplicity. We find that the best fit to the recent IceCube data in the $U_1$ case is obtained for $m_{U_1} = 710$\,GeV and $\chi^L_{de} = 1.25$ and results in a 9.5\% improvement over the SM fit. This is in contrast to the $S_1$ case, which for most of the parameter space considered in our work either results in a fit worse or comparable to the SM. Therefore we show in the left (right) plot of Fig.~\ref{fig:VLQ_S1} the contribution of $U_1$ ($S_1$) for mass of 710\,GeV and $\chi^L_{de} = 1.25$ ($y^L_{d \nu_e} = 1.25$), which gives $\delta = 9.5$\% ($\delta = -2.9$\%). All other couplings are set to zero. 

We see from the above analysis that no significant deviation from the SM prediction is seen in the current IceCube data. We use this information to put an upper bound on the $y^L_{d\nu_e} (\chi^L_{de})$ coupling as a function of $m_{S_1}$ ($m_{U_1}$) through a binned likelihood analysis with the Poisson likelihood function~\cite{Aartsen:2014muf}. This constraint obtained on the $S_1$ and $U_1$ parameter space is then compared with the results from the low-energy flavor experiments and the LHC in the next section.

 \section{Combined analysis of the low-energy flavor physics, LHC, and IceCube constraints}
 \label{sec:combined_analysis}

\begin{figure}[htb!]
\centering
\begin{subfigure}{.45\textwidth}
\includegraphics[width=0.9\linewidth,keepaspectratio]{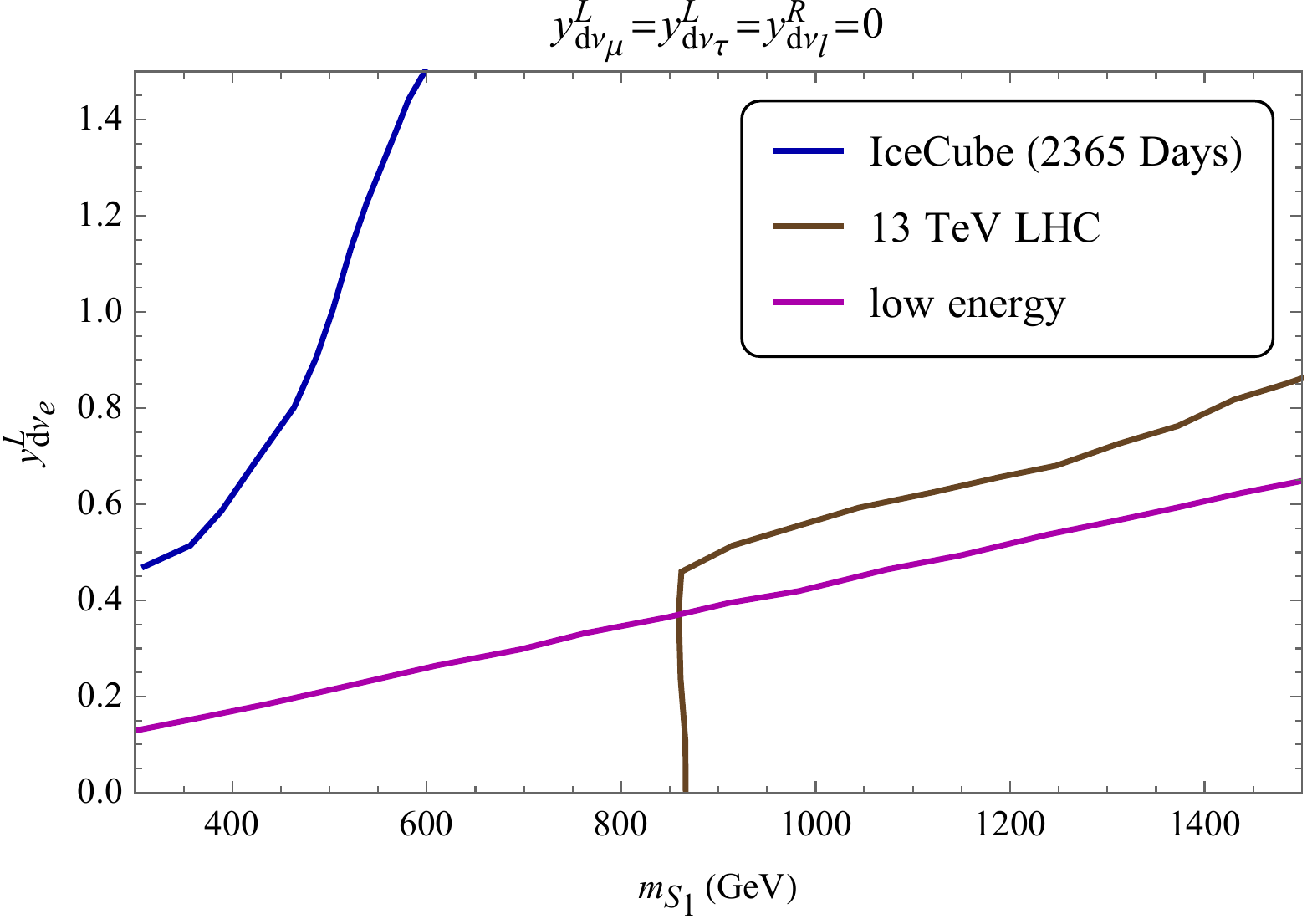}
\caption{}
\label{fig:SLQ1}
\end{subfigure}
\begin{subfigure}{.45\textwidth}
\includegraphics[width=0.9\linewidth, keepaspectratio]{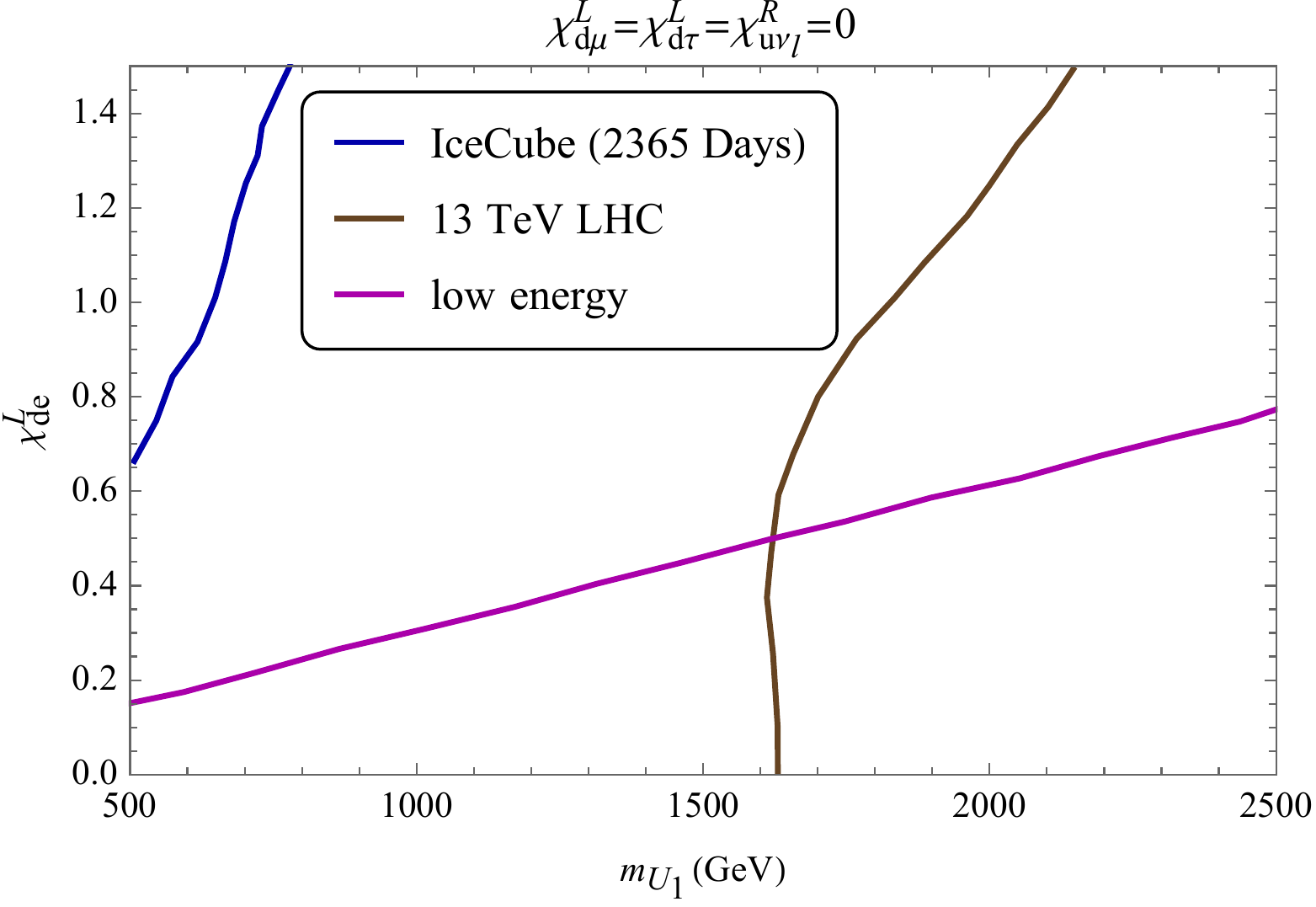}
\caption{}
\label{fig:VLQ1}
\end{subfigure}
\caption{The $m_{S_1}$-$y^L_{d\nu_e}$ (left panel) and $m_{U_1}$-$\chi^L_{de}$ (right panel) parameter space where the region above the blue line is disallowed at 2\,$\sigma$ level from 2635 days of IceCube data. We also show the low energy disallowed region (space above purple line) at the 2\,$\sigma$ level. The region to the left of the brown line is currently disallowed at 95\% C.L.\ by the 13\,TeV LHC data on the LQ direct searches via the $jj\nu\nu$, $jjee$, and $jj\mu\mu$ final states. The single LQ production is also included in the $S_1$ case.}
\label{fig:All}
\end{figure}

Our goal is to combine the low-energy flavor physics, LHC, and IceCube constraints on the parameter spaces associated with the $S_1$ and $U_1$ scenarios. The summary of our analysis of these constraints on the $m_{S_1}$($m_{U_1}$)-$y^L_{d\nu_e}(\chi^L_{de})$ parameter space is shown in Fig.~\ref{fig:All}. The LHC constraints from the LQ pair production with the dijet + MET, $jjee$, and $jj\mu\mu$ final states are considered and the currently allowed space, at 95\% C.L., is shown by the area below (right of) the brown line in case of $S_1$ ($U_1$). The region below 850\,GeV (1.6\,TeV) in the $S_1$ ($U_1$) case is completely excluded by the LHC data. The region below the purple line in Fig.~\ref{fig:All} is allowed by the flavor observables and the radiative decays of leptons. The region above the blue line is currently disallowed by IceCube data at 2\,$\sigma$ level. For our statistical analysis of the IceCube constraints, the LQ mass and the couplings are kept as free parameters, with $\gamma$ and $C_0$ fixed to the IceCube best fit data. We have used, for our numerical calculation, the ($1/3:1/3:1/3$) flavor ratio for the incoming flux. Even though the IceCube data alone favors $U_1$ over $S_1$ the actual parameter space allowed by the low-energy flavor experiments and LHC corresponds to the region where $S_1$ ($U_1$) marginally improves (spoils) the IceCube data fit when compared to the SM case. Overall we find that the limits obtained from the most recent IceCube data are considerably weaker when compared to the constraints from the low-energy observables and direct LQ searches at the LHC. This is mostly due to the current lack of statistics in the high-energy bins of the IceCube spectrum.

We next discuss the effects of inclusion of $y^R_{d\nu_\ell}$ on our analysis. The leading processes for the LQ pair production at the LHC will be via the initial state of $gg$, $u\bar{u}$, and $d\bar{d}$. The cross section in the $S_1$ case will be particularly enhanced through the $d\bar{d}$ initial state for large values of the right-handed $y_{d\nu_\ell}^R$ couplings. That regime will generate large branching ratio of $S_1$ to $j\nu$ and will, therefore, be strongly constrained by the $jj\nu\nu$ final state searches at the LHC. 
\begin{figure}[htb!]
\centering
\begin{subfigure}{.45\textwidth}
\includegraphics[width=0.9\linewidth, height=0.25\textheight,keepaspectratio]{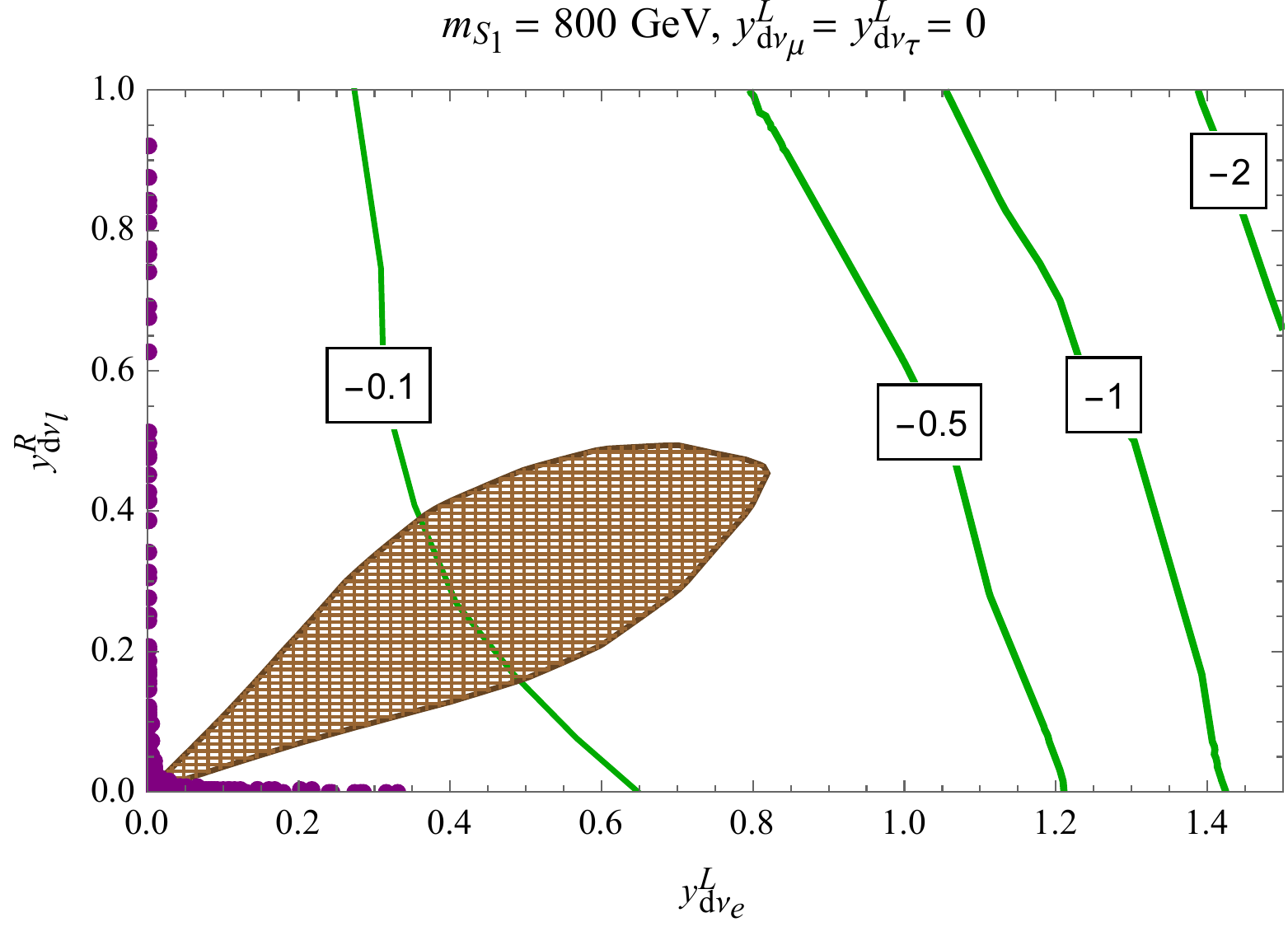}
\caption{}
\label{fig:rh1}
\end{subfigure}
\begin{subfigure}{.45\textwidth}
\includegraphics[width=0.9\linewidth,height=0.25\textheight,keepaspectratio]{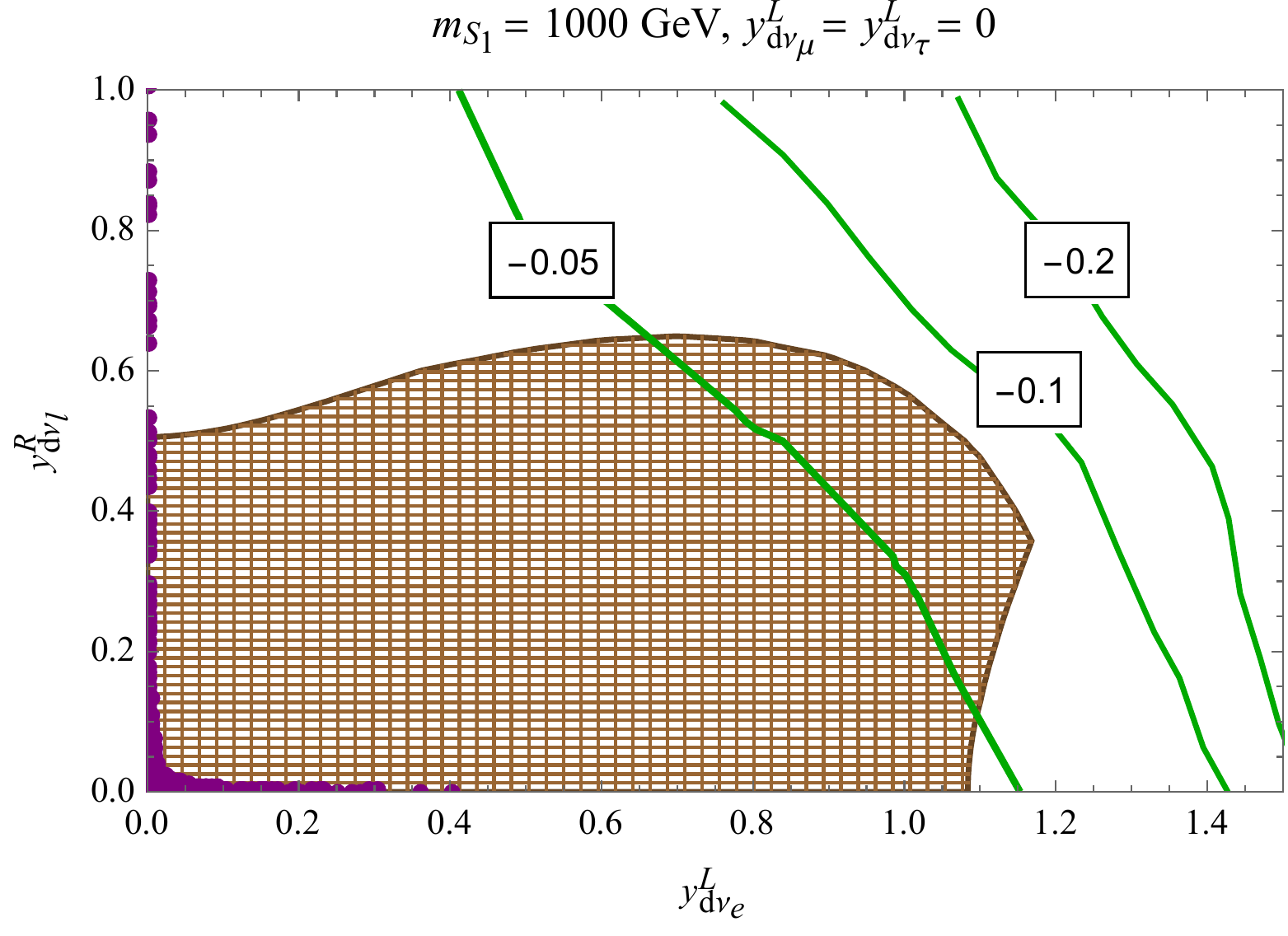}
\caption{}
\label{fig:rh2}
\end{subfigure}
\caption{The allowed parameter space at 95\% C.L.\ from the 13\,TeV LHC direct LQ searches is shown in brown in the $y^L_{d\nu_e}$-$y^R_{d\nu_\ell}$ plane, with $y^L_{d\nu_\mu} = y^L_{d\nu_\tau} = 0$ for $m_{S_1} = 800$\,GeV (left panel) and $m_{S_1} =1$\,TeV (right panel). The contours of constant value of $\delta$, as defined in Eq.~\eqref{eq:chi_sq}, show the percent change in $\chi^2$. The purple points are allowed at the 2\,$\sigma$ level by the low-energy flavor constraints discussed in Sec.~\ref{sec:low_energy}.}
\label{fig:limit_yR}
\end{figure}
We show in Fig.~\ref{fig:limit_yR} the allowed parameter space from the LHC in brown in the $y^L_{d\nu_e}$-$y^R_{d\nu_\ell}$ plane with $y^L_{d\nu_\mu} = y^L_{d\nu_\tau} = 0$ for $m_{S_1}$ = 800\,GeV (left panel) and 1\,TeV (right panel) and contours of constant values of $\delta$ in green. Since the IceCube data are more sensitive to the $\nu_e$ coupling we consider this particular choice to check the effect of $y^R_{d\nu_\ell}$ on $y^L_{d\nu_e}$. Clearly, the inclusion of $y^R_{d\nu_\ell}$ slightly spoils the fit of the IceCube data in the region of interest. This is expected since the coupling $y^R_{d\nu_\ell}$ only appears quadratically in the final state of the $\nu N$ cross section. Note that $y^R_{d\nu_\ell}$ is also tightly constrained by the flavor observables and we present in Fig.~\ref{fig:limit_yR} the allowed region by purple points. 

The inclusion of the $U_1$ couplings to the right-handed neutrinos follows the same pattern as in the $S_1$ case. Provided that the right-handed neutrinos do not contribute to the initial state, the current IceCube data are not sensitive to these couplings. The consideration of the right-handed neutrinos in the initial state will lead to a change in the initial flux at the source. This flux will depend on the mass of $\nu^R_\ell$ and also on the possible decay channels in case of heavy $\nu^R_\ell$. This is beyond the scope of present analysis. The resulting final high-energy cosmic neutrino flux ratios on earth by the possible mixing between the three active neutrinos and the fourth sterile neutrino have been studied in Ref.~\cite{Athar:2000yw}. The explanation of the PeV neutrinos at IceCube, with the consideration of heavy right-handed neutrino, acting as a dark matter has been studied in Ref.~\cite{Dev:2016qbd}.

\section{Conclusions}
\label{sec:conclusion}

We analyse the impact of the latest low-energy flavor physics measurements, LHC search limits, and IceCube data on the parameter space of the electroweak $SU(2)$ singlet scalar (vector) LQ $S_1$ ($U_1$). We perform, in particular, a thorough analysis of the viability of the $S_1$ ($U_1$) scenario assuming non-zero couplings between $S_1$ ($U_1$), down quark, and neutrinos (charged leptons) of all three generations, where the LQ in question couples to the SM lepton doublets. Consequentially, the $SU(2)$ symmetry of the SM requires that $S_1$ ($U_1$) couples up-type quarks to charged leptons (neutrinos).

We find that the limits obtained from the most recent IceCube data are considerably weaker when compared to the constraints from the low-energy observables and direct LQ searches at the LHC and we quantify this inferiority. We attribute this disparity in constraining power to the current lack of statistics in the high-energy bins of the IceCube spectrum. Even though the IceCube data alone favors $U_1$ over $S_1$, the viable  parameter space allowed by the low-energy flavor physics and the LHC data analyses singles out the region where $S_1$ ($U_1$) marginally improves (spoils) the IceCube data fit when compared to the SM case. In this region, in the $S_1$ ($U_1)$ case, $m_{S_1}\geq 900$\,GeV ($m_{U_1}\geq 1.6$\,TeV) and the relevant $S_1$-neutrino-down quark ($U_1$-neutrino-up quark) coupling is small. We have also verified that the couplings of both $S_1$ and $U_1$ to the right-handed neutrinos are not being sensitive to the current IceCube data provided that the right-handed neutrinos only contribute to the final state. Since we investigate scenarios where LQs primarily couple to the first generation quarks, the most important constraints originate from flavor physics measurements, followed by the LHC search limits that take over flavor physics limits in the large LQ mass regime.

\vspace{0.5cm}
{\bf Acknowledgments} 
We are grateful to Olcyr Sumensari, Boris Panes, and Damir Be\v cirevi\' c  for helpful discussions. M.P.\ would like to thank Anushree Ghosh for discussions on IceCube. S.F.\ and M.P.\ acknowledge support of the Slovenian Research Agency through research core funding No.\ P1-0035.

\appendix

\section{Formulas}\label{sec:appen}
We list here the different branching ratios used for our analysis in Sec.~\ref{sec:low_energy}.
\begin{eqnarray}
\mathrm{BR}(D^0 \rightarrow \mu^+ \mu^-) &=& \tau_{D}f_{D}^2 m_{D}^3 \frac{G_F^2}{64 \pi}\sqrt{1-\frac{4 m_\mu^2}{m_D^2}} \Bigg|\frac{m_\mu}{m_{D}}\frac{v^2}{m_{S_1}^2} (V^\ast y^L)_{12}(V^\ast y^L)_{22}\Bigg|^2
\end{eqnarray}
The $\pi \rightarrow \ell \bar{\nu}$ and the $\tau^- \rightarrow \pi^- \nu$ branching ratios at the leading order in SM are given by: 
\begin{align}
\label{eq:appSM}
&\mathrm{BR}(\pi \rightarrow \ell \bar{\nu})|_{\mathrm{SM}}\, =\,\tau_{\pi} \frac{G_F^2}{8\pi}f_\pi^2 m_\pi m_\ell^2  |V_{11}|^2 \Bigg(1-\frac{m_\ell^2}{m_\pi^2}\Bigg)^2,  \nonumber \\
&\mathrm{BR}(\tau^- \rightarrow \pi^- \nu)|_{\mathrm{SM}} \,=\,\tau_{\tau} \frac{G_F^2}{16\pi}f_\pi^2 m_\tau^3 |V_{11}|^2  \Bigg(1-\frac{m_\pi^2}{m_\tau^2}\Bigg)^2.
\end{align}
The electroweak corrections to $\mathrm{BR}(\pi \rightarrow \ell \bar{\nu})$ were calculated in Ref. \cite{Cirigliano:2007xi} and for $\mathrm{BR}(\tau^- \rightarrow \pi^- \nu)$ in Ref.  \cite{Erler:2002mv}. 
The relevant branching ratios in the LQ models is given by,
\begin{align}
\label{eq:app}
\mathrm{BR}(\pi \rightarrow \ell \bar{\nu}) &=\tau_{\pi} \frac{G_F^2}{8\pi}f_\pi^2 m_\pi^3 \Bigg(1-\frac{m_\ell^2}{m_\pi^2}\Bigg)^2\left[\frac{m_\ell^2}{m_\pi^2} |V_{11}|^2  + \frac{m_\ell^2}{m_\pi^2}\frac{2v^2}{C m_{\mathrm{LQ}}^2} {\mathrm{Re}}\left(  V^\ast_{11} (y^{L}_{q\ell})_{1j}\sum_{i=1}^3 U_{ji}^\ast (y^L_{q\nu})_{1i}\right) \right. \nonumber \\
 &  + \left.  \frac{v^2}{C m_{\mathrm{LQ}}^2} \left(\frac{m_\ell}{m_\pi}(y^{L}_{q\ell})_{1j}\sum_{i=1}^3 |(y^L_{q\nu})_{1i}|^2 +C' \sum_{i=1}^3 |y^R_{1i}|^2 m_\pi^2 \Bigg|\frac{(y^L_{q\ell})_{1j}}{m_u + m_d}\Bigg|^2\right)\right],  \nonumber \\
\mathrm{BR}(\tau^- \rightarrow \pi^- \nu) &=\tau_{\tau} \frac{G_F^2}{16\pi}f_\pi^2 m_\pi^2 m_\tau \Bigg(1-\frac{m_\pi^2}{m_\tau^2}\Bigg)^2\left[\frac{m_\tau^2}{m_\pi^2} |V_{11}|^2  + \frac{m_\tau^2}{m_\pi^2}\frac{v^2}{C m_{\mathrm{LQ}}^2} {\mathrm{Re}}\left(  V^\ast_{11} (y^L_{q\ell})_{13}\sum_{i=1}^3 U_{3i}^\ast (y^L_{q\nu})_{1i}\right) \right. \nonumber \\
&+ \left.  \frac{v^2}{C m_{\mathrm{LQ}}^2} \left(\frac{m_\tau}{m_\pi}(y^{L}_{q\ell})_{13}\sum_{i=1}^3 |(y^L_{q\nu})_{1i}|^2 +C' \sum_{i=1}^3 |y^R_{1i}|^2 m_\pi^2 \Bigg|\frac{(y^L_{q\ell})_{13}}{m_u + m_d}\Bigg|^2\right)\right],
\end{align}
with $y^{L}_{q\nu} = y^L U (V^\dag \chi^L U)$, $y^{L}_{q\ell} = V^\ast y^L (\chi^L)$, $C = 4 (2)$, and $C' = 1 (2)$ in the $S_1 (U_1)$ case. The light quark masses are determined at the LQ mass scale. The subscript $j$ in Eq.~\eqref{eq:app} takes on the values of 1 and 2 for $e^-$ and $\mu^-$ respectively.

\section{Neutrino-nucleon differential cross sections}
\label{sec:2a}

The neutrino-nucleon scattering in the SM gives rise to the charged current ($\nu_\ell N \rightarrow \ell  X$) and the neutral current ($\nu_\ell N \rightarrow \nu_\ell X$) interactions mediated by $W$ and $Z$ bosons, respectively. The target nucleon $N$ is an isoscalar nucleon with $ N =(n+p)/2$, $X$ is the hadronic final state, and $\ell = e,\mu,\tau$. The SM differential cross sections in terms of the scaling variables are given as 
\begin{eqnarray}\label{eq:totCS_SM}
\frac{d^2\sigma_{\nu N}^{\rm CC}}{dxdy} &=& \frac{2G_F^2 m_N E_\nu}{\pi}
\left(\frac{m_W^2}{Q^2+m_W^2}\right)^2 \left[xq(x,Q^2)+x\bar{q}(x,Q^2)(1-y)^2\right],\label{cc}\\
\frac{d^2\sigma_{\nu N}^{\rm NC}}{dxdy} &=& \frac{G_F^2m_N E_\nu}{2\pi}
 \left[xq^0(x,Q^2)+x\bar{q}^0(x,Q^2)(1-y)^2\right],\label{nc}
\end{eqnarray}
where $-Q^2$ is the invariant momentum-square transfer to the exchanged vector boson, $m_N$ and $m_{W(Z)}$ are the nucleon and intermediate $W(Z)$ boson masses, respectively, and $G_F$ = 1.166378 $\times 10^{-5}$\,GeV$^{-2}$ is the Fermi coupling constant. The differential distributions in Eqs.~\eqref{cc} and \eqref{nc} are with respect to the Bjorken scaling variable $x$ and the inelasticity parameter $y$, where 
\begin{eqnarray}
x=\frac{Q^2}{2m_N yE_\nu}~~~{\rm and}~~~y=\frac{E_\nu-E_\ell}{E_\nu}\label{xy}.
\end{eqnarray}
$E_\ell$ denotes the energy carried away by the outgoing lepton or the neutrino in the laboratory frame while $x$ is the fraction of the initial nucleon momentum taken by the struck quark. Here, $q(x,Q^2)$ and $\bar{q}(x,Q^2)$ ($q^0(x,Q^2)$ and $\bar{q}^0(x,Q^2)$) are the quark and anti-quark density distributions in a proton, respectively, summed over valence and sea quarks of all flavors relevant for CC (NC) interactions:  
\begin{align}
q(x,Q^2) &= \frac{u_v(x,Q^2)+d_v(x,Q^2)}{2}+\frac{u_s(x,Q^2)+d_s(x,Q^2)}{2}+s_s(x,Q^2)+b_s(x,Q^2),\label{eq:QQ0bar1}\\
\bar{q}(x,Q^2) &= \frac{u_s(x,Q^2)+d_s(x,Q^2)}{2}+c_s(x,Q^2)+t_s(x,Q^2), \label{eq:QQ0bar2} \\
q^0(x,Q^2) &=\left(\frac{m_Z^2}{Q^2+m_Z^2}\right)^2\Bigg[\left(\frac{u_v(x,Q^2)+d_v(x,Q^2)}{2}+\frac{u_s(x,Q^2)+d_s(x,Q^2)}{2}\right)(L_u^2+L_d^2) \nonumber\\ 
&  +\frac{u_s(x,Q^2)+d_s(x,Q^2)}{2}(R_u^2+R_d^2)+(s_s(x,Q^2)+b_s(x,Q^2))(L_d^2+R_d^2) \nonumber\\
& +(c_s(x,Q^2)+t_s(x,Q^2))(L_u^2+R_u^2)\Bigg],\label{eq:QQ0bar3}\\ 
\bar{q}^0(x,Q^2) &=\left(\frac{m_Z^2}{Q^2+m_Z^2}\right)^2\Bigg[ \left(\frac{u_v(x,Q^2)+d_v(x,Q^2)}{2}+\frac{u_s(x,Q^2)+d_s(x,Q^2)}{2}\right)(R_u^2+R_d^2) \nonumber\\ 
&+\frac{u_s(x,Q^2)+d_s(x,Q^2)}{2}(L_u^2+L_d^2)+(s_s(x,Q^2)+b_s(x,Q^2))(L_d^2+R_d^2) \nonumber\\
& +(c_s(x,Q^2)+t_s(x,Q^2))(L_u^2+R_u^2) \Bigg]\label{eq:QQ0bar4}, 
\end{align}
with the chiral couplings given by $L_u=1-(4/3)x_W,~L_d=-1+(2/3)x_W,~R_u=-(4/3)x_W$, and $R_d=(2/3)x_W$, where $x_W=\sin^2\theta_W$ and $\theta_W$ is the weak mixing angle. For the $\bar{\nu} N$ cross sections Eqs.~\eqref{cc} and \eqref{nc} are the same but with each quark distribution function replaced by the corresponding anti-quark distribution function, and vice-versa, i.e., $q(x,Q^2)\leftrightarrow \bar{q}(x,Q^2),~q^0(x,Q^2)\leftrightarrow \bar{q}^0(x,Q^2)$. The parton distribution functions (PDFs) of the quarks are evaluated at energy $Q^2$, and the Mathematica package MSTW~\cite{Martin:2009iq} is used throughout this work.

There are also neutrino-electron interactions, but they can be generally neglected with respect to the neutrino-nucleon cross section because of the smallness of electron's mass, except for the resonant formation of the intermediate $W^-$ boson in the $\bar{\nu}_e e$ interactions at around  $E_\nu = m_W^2/(2 m_e) = 6.3\times 10^6$\,GeV, known as the Glashow resonance. The differential cross sections for all the neutrino electron reactions are listed in Ref.~\cite{Gandhi:1995tf}. 

\subsection{$\nu_j N$ cross sections in the presence of the $S_1$} \label{app_2a}
The differential $\nu_j N$ cross sections in the presence of the $S_1$ interaction are given by
\begin{eqnarray}\label{eq:s1_cs}
\frac{d^2\sigma_{\nu_j N}^{ch}}{dx dy} &=&\frac{m_N E_\nu}{16 \pi}|(y^{L}U)_{1j}|^2 \mathcal{N}^{ch} \Bigg(\frac{1}{|2 x m_N E_\nu-m_{S_1}^2+i \Gamma_{S_1}m_{S_1}|^2}\Bigg[\frac{u_v+d_v}{2}+\frac{u_s+d_s}{2}\Bigg] \nonumber \\
&+&\frac{1}{(Q^2-2 x m_N E_\nu-m_{S_1}^2)^2}(1-y)^2 \frac{u_s+d_s}{2} \Bigg),  
\end{eqnarray}
where $j$ = 1,2,3 with $ch$ = $CC,~NC$. The coefficients are given by $\mathcal{N}^{CC}=\sum_{k=1}^3|(V^*y^{L})_{1k}|^2$, $\mathcal{N}^{NC}=\sum_{k=1}^3|(y^LU)_{1k}|^2 +\sum_{k=1}^3  |y^{R}_{1k}|^2$. The decay width $\Gamma_{S_1}$ of $S_1$ given by
 \begin{eqnarray}
 \Gamma_{S_1} &=& \frac{m_{S_1}}{16\pi}\Bigg[ \sum_{i=1}^3 |(y^{L}U)_{1i}|^2 + \sum_{i,j = 1}^3  |(V^*y^{L})_{ij}|^2   + \sum_{i=1}^3  |y^{R}_{1i}|^2\Bigg].
 \end{eqnarray}
Note that the effect of the right-handed couplings to the neutrinos is only visible in NC interactions.

\subsection{$\nu_j N$ cross sections in the presence of the $U_1$} \label{app_2b}

The modified $q^0(x,Q^2)$ and $\bar{q}^0(x,Q^2)$ listed in Eqs.~\eqref{eq:QQ0bar3} and~\eqref{eq:QQ0bar4}, in the presence of $U_1$, are given below.  
\begin{dmath}\label{eq:VLQQQ0bar1}
q^0(x,Q^2) = \left(\frac{m_Z^2}{Q^2+m_Z^2}\right)^2\Bigg[\left(\frac{u_v(x,Q^2)+d_v(x,Q^2)}{2}+\frac{u_s(x,Q^2)+d_s(x,Q^2)}{2}\right)L_d^2 + \frac{u_s(x,Q^2)+d_s(x,Q^2)}{2} 
(R_u^2+R_d^2)  \Bigg] 
+ \left[\frac{u_v(x,Q^2)+d_v(x,Q^2)}{2}+\frac{u_s(x,Q^2)+d_s(x,Q^2)}{2}  \right]
\left| \frac{m_Z^2}{Q^2+m_Z^2} L_u +  \frac{|(V^\dagger\chi^LU)_{1i}|^2}{2\sqrt{2}G_F}\frac{1}{Q^2- 2 x M_N E_\nu - m_{U_1}^2} \right|^2 \end{dmath}
\begin{dmath}\label{eq:VLQQQ0bar2}
\bar{q}^0(x,Q^2) =\left(\frac{m_Z^2}{Q^2+m_Z^2}\right)^2\Bigg[ \left(\frac{u_v(x,Q^2)+d_v(x,Q^2)}{2}+\frac{u_s(x,Q^2)+d_s(x,Q^2)}{2}\right)(R_u^2+R_d^2) +\frac{u_s(x,Q^2)+d_s(x,Q^2)}{2}L_d^2\Bigg]+\frac{u_s(x,Q^2)+d_s(x,Q^2)}{2} \left|\frac{m_Z^2}{Q^2+m_Z^2} L_u +  \frac{|(V^\dagger\chi^LU)_{1j}|^2}{2\sqrt{2}G_F}\frac{1}{2 x m_N E_\nu - m_{U_1}^2 +i \Gamma_{U_1} m_{U_1}}\right|^2
\end{dmath}
where $j = 1,2,3$.
Eqs.~\eqref{eq:VLQQQ0bar1} and~\eqref{eq:VLQQQ0bar2} correspond to the case when the same flavor neutrino is in the initial and the final states and there is an interference with the SM contribution. There will be additional contributions from the cases where the final state will consist of the right-handed neutrinos or will have a neutrino of different flavor from the initial one and are given by
\begin{eqnarray}
\frac{d^2\sigma_{\nu_j N}^{NC}}{dx dy}&=&\frac{E_\nu}{8 G_F^2}|(V^\dagger\chi^LU)_{1j}|^2\Bigg(\frac{1}{(Q^2-2x m_N E_\nu-m_{U_1}^2)^2}\left(y^2 \sum_{k=1}^3 |\chi^R_{1k}|^2  + \sum_{\substack{k=1\\ k\neq j}}^3|(V^\dagger\chi^LU)_{1k}|^2\right)\nonumber \\
&&\Bigg[\frac{u_v+d_v}{2}+\frac{u_s+d_s}{2}\Bigg]+(1-y)^2 \left|\frac{1}{2 x m_N E_\nu - m_{U_1}^2 +i\Gamma_{U_1} m_{U_1}}\right|^2 \sum_{\substack{k=1\\ k\neq j}}^3|(V^\dagger\chi^LU)_{1k}|^2 \frac{u_s+d_s}{2} \nonumber \Bigg).
\end{eqnarray}
In case of CC interactions, the coefficient $m_W^2/(Q^2+m_W^2)$ in Eq.~\eqref{eq:totCS_SM} is modified to 
\begin{eqnarray}
\left(\frac{m_W^2}{(Q^2+m_W^2)}\right)^2q(x,Q^2) \Rightarrow  \left|\frac{m_W^2}{(Q^2+m_W^2)}+\frac{\chi^L_{dj} (V^\dagger\chi^LU)_{1j}}{2\sqrt{2}G_F} \frac{1}{Q^2- 2 x m_N E_\nu - m_{U_1}^2}\right|^2 q(x,Q^2) \nonumber \\
\left(\frac{m_W^2}{(Q^2+m_W^2)}\right)^2 \bar{q}(x,Q^2) \Rightarrow  \left|\frac{m_W^2}{(Q^2+m_W^2)}+\frac{\chi^L_{dj} (V^\dagger\chi^LU)_{1j}}{2\sqrt{2}G_F} \frac{1}{2 x m_N E_\nu - m_{U_1}^2 +i \Gamma_{U_1} m_{U_1}}\right|^2 \bar{q}(x,Q^2), 
\end{eqnarray}
taking into account the interference terms only when the final state is similar to the SM. The decay width of $U_1$ is given by 
\begin{eqnarray}
\Gamma_{U_1}&=&\frac{m_{U_1}}{24\pi}\left[\sum_{i=1}^3 |\chi^L_{1i}|^2+\sum_{i=1}^3|(V^\dagger\chi^LU)_{1i}|^2+\sum_{i=1}^3 |\chi^R_{1i}|^2\right].
\end{eqnarray}

\end{document}